\documentclass[showpacs,preprintnumbers,amsmath,amssymb]{revtex4}

\usepackage{graphicx}
\usepackage{dcolumn}
\usepackage{bm}

\newcommand{\bq}{\begin{equation}}
\newcommand{\eq}{\end{equation}}
\newcommand{\bqa}{\begin{eqnarray}}
\newcommand{\eqa}{\end{eqnarray}}
\newcommand{\nn}{\nonumber \\}

\def\be     {\begin{equation}}
\def\ee     {\end{equation}}
\def\bea        {\begin{eqnarray}}
\def\eea        {\end{eqnarray}}
\def\bnn    {\begin{eqnarray*}}
\def\enn    {\end{eqnarray*}}

\begin{document}

\title{Emergence of localized magnetic moments near antiferromagnetic quantum criticality}
\author{Ki-Seok Kim}
\affiliation{Department of Physics, POSTECH, Pohang, Gyeongbuk 790-784, Korea \\ Institute of Edge of Theoretical Science (IES), POSTECH, Pohang, Gyeongbuk 790-784, Korea}
\date{\today}

\begin{abstract}
We revisit an antiferromagnetic quantum phase transition with $\bm{Q} = 2 \bm{k}_{F}$, where $\bm{Q}$ is an ordering wave vector and $\bm{k}_{F}$ is a Fermi momentum. Reformulating the Hertz-Moriya-Millis theory within the strong coupling approach to diagonalize the spin-fermion coupling term and performing the scaling analysis for an effective field theory with quantum corrections in the Eliashberg approximation, we propose a novel interacting fixed point for this antiferromagnetic quantum phase transition, where antiferromagnetic spin fluctuations become locally critical to interact with renormalized electrons. The emergence of local quantum criticality suggests a mechanism of $\omega/T$ scaling for antiferromagnetic quantum criticality, generally forbidden in the context of the Hertz-Moriya-Millis theory.
\end{abstract}


\maketitle

\section{Introduction}

A standard theoretical framework for quantum phase transitions in ``good" metals is Hertz-Moriya-Millis theory, describing non-Fermi liquid physics near quantum criticality in terms of local order-parameter fluctuations \cite{HMM}. An essential aspect of this field-theory approach is that it does not allow the frequency ($\omega$) over temperature ($T$) scaling physics around quantum criticality, which originates from the existence of abundant soft modes near the Fermi surface, giving rise to the fact that the Hertz-Moriya-Millis theory lives above its upper critical dimension and breaks the hyperscaling relation \cite{QCP_Review1,QCP_Review2}. It has been pointed out that the structure of the Hertz-Moriya-Millis theory may be modified, where the interaction parameter in the quartic term of the Hertz-Moriya-Millis theory is not a constant but a complicated function for frequency and momentum, giving rise to nonlocal correlations effectively, if one goes beyond the Eliashberg approximation \cite{BVK,Pepin_FM,Chubukov_FM,Chubukov_AFM,Metlitski_SDW}. Unfortunately, the role of such nonlocal interactions remains inconclusive.

Recently, the possibility of $\omega/T$ scaling has been addressed in a study on heavy-fermion quantum criticality \cite{HFQCP_NLsM_RG}. Based on the nonlinear $\sigma-$model description for dynamics of localized spins, Kondo fluctuations are shown to cause nonlocal interactions between such spin fluctuations. Performing the renormalization group analysis, nonlocal interactions between spin-wave modes turn out to make the spin-wave velocity vanish logarithmically at the quantum critical point. Local quantum criticality \cite{Si_LQCP} driven by nonlocal interactions between order-parameter fluctuations will allow the $\omega/T$ scaling physics in dynamic response functions.

In this study, we revisit an antiferromagnetic quantum phase transition in the system of itinerant electrons with Fermi-surface nesting. In particular, we focus on the case of $\bm{Q} = 2 \bm{k}_{F}$, where $\bm{Q}$ is an ordering wave vector and $\bm{k}_{F}$ is a Fermi momentum. An idea is to take a strong coupling approach in the Hertz-Moriya-Millis theory \cite{With_MDKim}, which diagonalizes the spin-fermion coupling term, reformulating the Hertz-Moriya-Millis theory in terms of fermionic holons (renormalized electrons) and bosonic spinons (directional spin fluctuations) \cite{FMLQCP}, referred to as U(1) slave spin-rotor theory (section II).  As a result, we obtain an effective field theory in the two-patch construction \cite{SSL_U1GT}, describing holons, longitudinal (amplitude) antiferromagnetic (critical) and ferromagnetic (gapped) fluctuations, transverse spin fluctuations (spinons), spin singlet (gauge) fluctuations (gapless), and their interactions (section III-A). Introducing quantum corrections into the effective field theory within the Eliashberg approximation (section III-A), we perform the scaling analysis at the ``tree" level. Considering the standard scaling analysis \cite{SSL_U1GT,U1GT_Scaling} in this renormalized effective theory, we fail to find a critical field theory, where interaction vertices turn out to be relevant (section III-C). On the other hand, assuming that the ``transverse" momentum along the Fermi surface do not change under the scale transformation, we find a critical field theory in terms of renormalized electrons and transverse spin fluctuations, where dynamics of transverse spin fluctuations becomes locally critical (section III-D). This local quantum criticality suggests a mechanism of $\omega/T$ scaling, generally forbidden in the context of the Hertz-Moriya-Millis theory, since this fixed point is interacting in nature, distinguished from that of the Hertz-Moriya-Millis theory. Although we focus on the case of two dimensions, we find the same critical field theory in three dimensions. Emergence of localized magnetic moments is the nature of this novel fixed point in the antiferromagnetic quantum phase transition with $\bm{Q} = 2 \bm{k}_{F}$.


\begin{figure}[t]
\includegraphics[width=0.8\textwidth]{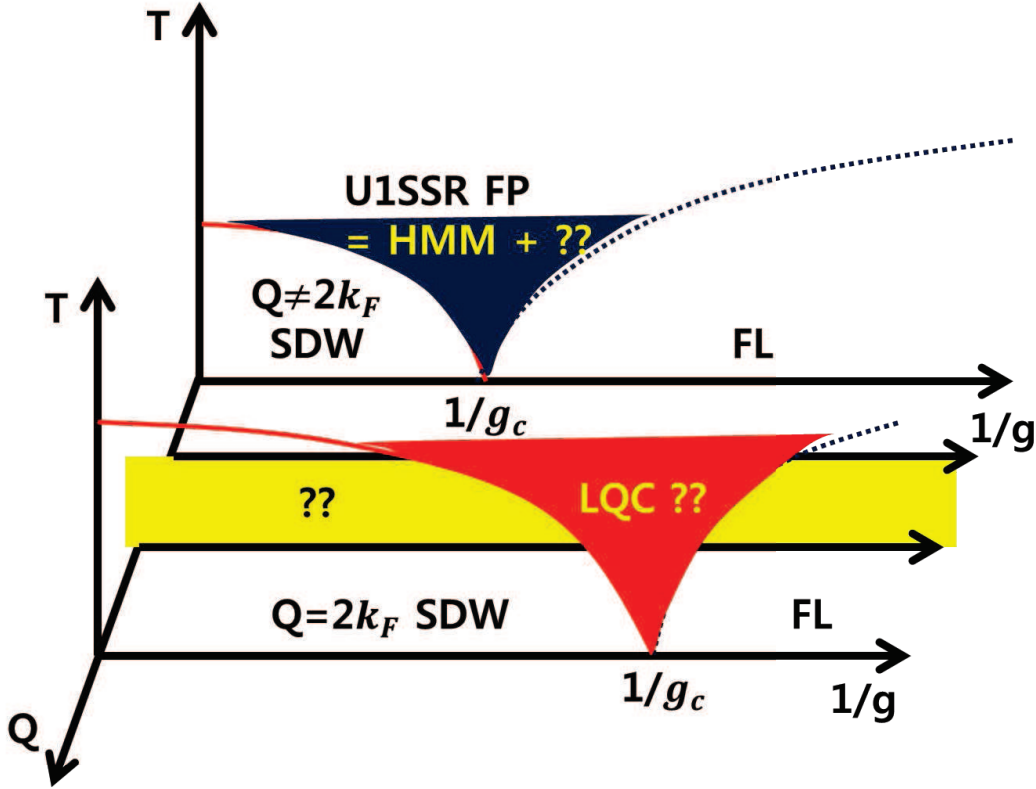}
\caption{A schematic phase diagram for antiferromagnetic quantum phase transitions. The $\bm{x}-$axis is the inverse of a spin-fermion coupling constant, and the $\bm{y}-$axis is temperature, constituting a conventional phase diagram. Here, we suggest to add an additional axis into this phase diagram, which corresponds to an ordering wave-vector, but changeable appropriately to be related with Fermi-surface nesting. ``SDW" and ``FL" denote spin density wave and Fermi liquid, respectively. ``U1SSR FP" represents U(1) slave spin-rotor fixed point, and ``HMM" and ``LQC" express Hertz-Moriya-Millis quantum criticality and local quantum criticality, respectively. It is a conventional view that the fixed point of an antiferromagnetic quantum critical point with $\bm{Q} \not= 2 \bm{k}_{F}$ is described by the Hertz-Moriya-Millis theory. Applying the U(1) slave spin-rotor formulation to this antiferromagnetic quantum phase transition, regarded to be a strong coupling approach, we find possibility for the emergence of another fixed point, where the Hertz-Moriya-Millis theory becomes modified to contain relevant interactions between transverse spin fluctuations and renormalized electrons (section III-B). It is not possible to clarify the condition for which fixed point will be realized between the the Hertz-Moriya-Millis and U(1) slave spin-rotor fixed points at present. An essential aspect of the present study is that the structure of the Hertz-Moriya-Millis theory breaks down at an antiferromagnetic quantum critical point with $\bm{Q} = 2 \bm{k}_{F}$, where interactions between longitudinal spin fluctuations and renormalized electrons, referred to as the spin-fermion coupling usually, turn out to be irrelevant (sections III-C and III-D). This situation differs from the case with $\bm{Q} \not= 2 \bm{k}_{F}$, where the structure of the Hertz-Moriya-Millis theory is preserved at least partially for the U(1) slave spin-rotor fixed point. The role of transverse spin fluctuations becomes more dominant in the antiferromagnetic quantum criticality with $\bm{Q} = 2 \bm{k}_{F}$, giving rise to a novel critical field theory within the U(1) slave spin-rotor formulation, where renormalized electrons interact with localized transverse spin fluctuations (section III-D).} \label{Phase_Diagram}
\end{figure}

A view point of the present study can be described by a schematic phase diagram of Fig. 1, where the $\bm{x}-$axis is the inverse of a spin-fermion coupling constant and the $\bm{y}-$axis is temperature, constituting a conventional phase diagram. We suggest to add an additional axis to this phase diagram, which corresponds to an ordering wave-vector, but changeable appropriately to be related with Fermi-surface nesting. It is a conventional view that the fixed point of an antiferromagnetic quantum critical point with $\bm{Q} \not= 2 \bm{k}_{F}$ is described by the Hertz-Moriya-Millis theory. Applying the strong coupling approach above to the antiferromagnetic quantum phase transition with $\bm{Q} \not= 2 \bm{k}_{F}$, we check the possibility of a novel fixed point beyond the Hertz-Moriya-Millis theory (section III-B). Although the Hertz-Moriya-Millis theory is a critical field theory for this antiferromagnetic quantum critical point, the U(1) slave spin-rotor formulation allows us to reach another fixed point, for which interactions between transverse spin fluctuations and renormalized electrons are responsible. As a result, the Hertz-Moriya-Millis theory becomes modified to contain such additional relevant interactions. It is not possible to clarify the condition for which fixed point will be realized between the the Hertz-Moriya-Millis and U(1) slave spin-rotor fixed points at present. Our critical observation is that the Hertz-Moriya-Millis theory is not invariant under the renormalization group transformation when the ordering wave vector is near $\bm{Q} = 2 \bm{k}_{F}$ (sections III-C and III-D). This situation differs from the case with $\bm{Q} \not= 2 \bm{k}_{F}$, where the structure of the Hertz-Moriya-Millis theory is preserved at least partially for the U(1) slave spin-rotor fixed point. Interactions between longitudinal spin fluctuations and renormalized electrons, referred to as the spin-fermion coupling usually, turn out to be irrelevant, responsible for breakdown of the Hertz-Moriya-Millis theory (sections III-C and III-D). Instead, the role of transverse spin fluctuations becomes more dominant in the antiferromagnetic quantum criticality with $\bm{Q} = 2 \bm{k}_{F}$, giving rise to a novel critical field theory within the U(1) slave spin-rotor formulation, where renormalized electrons interact with localized transverse spin fluctuations (section III-D), as discussed above.

\section{U(1) slave spin-rotor theory}

\subsection{A minimal model: Two patch construction}

We start from the Hubbard model \bqa && H = - t \sum_{ij} (c_{i\sigma}^{\dagger} c_{j\sigma} + H.c.) + U \sum_{i} n_{i\uparrow} n_{i\downarrow} , \eqa where $c_{i\sigma}$ represents an electron field with spin $\sigma$ at site $i$, and $t$ and $U$ are hopping and interaction parameters, respectively.

Focusing on magnetic instability, we perform the Hubbard-Stratonovich transformation for the spin-triplet channel \bqa && Z = \int D c_{i\sigma} D \boldsymbol{\Phi}_{i} \exp\Bigl[ - \int_{0}^{\beta} d \tau \Bigl\{ \sum_{i} c_{i\sigma}^{\dagger} (\partial_{\tau} - \mu) c_{i\sigma} - t \sum_{ij} (c_{i\sigma}^{\dagger} c_{j\sigma} + H.c.) - \sum_{i} c_{i\alpha}^{\dagger} \boldsymbol{\Phi}_{i} \cdot \boldsymbol{\sigma}_{\alpha\beta} c_{i\beta} + \frac{1}{2g} \sum_{i} \boldsymbol{\Phi}_{i}^{2} \Bigr\} \Bigr] , \eqa where $\boldsymbol{\Phi}_{i}$ is an order parameter of magnetization at site $i$. $g$ is an interaction parameter for the triplet channel, proportional to $U$, and $\mu$ is a chemical potential.

In this study we consider an antiferromagnetic transition, where the ordering wave vector $\bm{Q}$ is given by twice of the Fermi momentum $\bm{k}_{F}$, i.e., $\bm{Q} = 2 \bm{k}_{F}$. On the other hand, one may consider the case of $\bm{Q} \not= 2 \bm{k}_{F}$, where the structure of a Fermi surface is shown in Ref. \cite{Metlitski_SDW}, for example, associated with high T$_{c}$ cuprates. The key difference between these two cases lies in the scale transformation for longitudinal and transverse momenta, orthogonal to and along the Fermi surface, respectively. We discuss this issue in the next section.

We write down an order parameter field as follows \bqa && \boldsymbol{\Phi}_{i} \cdot \boldsymbol{\sigma}_{\alpha\beta} = (e^{i \boldsymbol{Q} \cdot \boldsymbol{r}_{i}} m + \delta m_{i}) \boldsymbol{n}_{i} \cdot \boldsymbol{\sigma}_{\alpha\beta} , \eqa where $m$ is an antiferromagnetic order parameter with a nesting vector $\bm{Q}$ of the Fermi surface, $\delta m_{i}$ is an amplitude-fluctuation field, and $\bm{n}_{i}$ is a directional-fluctuation field at site $i$.

Inserting this expression into the partition function, we obtain \bqa && Z = \int D c_{i\sigma} D \delta m_{i} D \bm{n}_{i} \delta(|\bm{n}_{i}|^{2} - 1) \exp\Bigl[ - \int_{0}^{\beta} d \tau \Bigl\{ \sum_{i} c_{i\sigma}^{\dagger} (\partial_{\tau} - \mu) c_{i\sigma} - t \sum_{ij} (c_{i\sigma}^{\dagger} c_{j\sigma} + H.c.) \nn && - \sum_{i} (e^{i \boldsymbol{Q} \cdot \boldsymbol{r}_{i}} m + \delta m_{i}) c_{i\alpha}^{\dagger} \boldsymbol{n}_{i} \cdot \boldsymbol{\sigma}_{\alpha\beta} c_{i\beta} + \frac{1}{2g} \sum_{i} (e^{i \boldsymbol{Q} \cdot \boldsymbol{r}_{i}} m + \delta m_{i})^{2} \Bigr\} \Bigr] . \eqa It is straightforward to perform the Fourier transformation, given by \bqa && Z = \int D c_{\bm{k}\sigma} D \delta m_{\bm{k}'} D \bm{n}_{\bm{q}} \delta(\sum_{\bm{q}} \bm{n}_{\bm{q}} \bm{n}_{-\bm{q}} - 1) \exp\Bigl[ - \int_{0}^{\beta} d \tau \Bigl\{ \sum_{\bm{k}} c_{\bm{k}\sigma}^{\dagger} (\partial_{\tau} - \mu + \epsilon_{\bm{k}}) c_{\bm{k}\sigma} \nn && - \sum_{\bm{k}} \frac{1}{L^{d}} \sum_{\bm{q}} m c_{\bm{k}\alpha}^{\dagger} \boldsymbol{n}_{\bm{q}} \cdot \boldsymbol{\sigma}_{\alpha\beta} c_{\bm{k}+\bm{Q}+\bm{q}\beta} - \sum_{\bm{k}} \frac{1}{L^{d}} \sum_{\bm{k}'} \frac{1}{L^{d}} \sum_{\bm{q}}  \delta m_{\bm{k}'} c_{\bm{k}\alpha}^{\dagger} \boldsymbol{n}_{\bm{q}} \cdot \boldsymbol{\sigma}_{\alpha\beta} c_{\bm{k}+\bm{k}'+\bm{q}\beta} \nn && + \frac{1}{2g} \sum_{\bm{q}} \Bigl( \delta m_{\bm{q}} \delta m_{-\bm{q}} + m [\delta(\bm{q} - \bm{Q}) + \delta(\bm{q} + \bm{Q})] \delta m_{\bm{q}} \Bigr) + L^{d} \frac{1}{2g} m^{2} \Bigr\} \Bigr] , \eqa where $L^{d}$ is a volume of the system. Since $c_{\bm{k}\sigma}$ is coupled to $c_{\bm{k}+\bm{Q}\sigma}$, it is natural to introduce a spinor $\bm{c}_{\bm{k}\sigma} = \left(\begin{array}{c} c_{\bm{k}\sigma} \\ c_{\bm{k}+\bm{Q}\sigma} \end{array} \right)$ in the momentum space. Performing the Fourier transformation for this spinor, we obtain $\bm{c}_{i\sigma} = \left(\begin{array}{c} c_{i+\sigma} \\ c_{i-\sigma} \end{array} \right)$, where $c_{i+(-)\sigma}$ is an electron field near the $+$ ($-$) patch of the Fermi surface. In the same way we introduce $\delta \bm{m}_{\bm{k}'} = \left(\begin{array}{c} \delta m_{\bm{k}'} \\\delta m_{\bm{k}'+\bm{Q}} \end{array} \right)$ and obtain $\delta \bm{m}_{i} = \left(\begin{array}{c} \delta m_{i1} \\\delta m_{i2} \end{array} \right)$ after the Fourier transformation, where $\delta m_{i1}$ and $\delta m_{i2}$ represent ``ferromagnetic" and ``antiferromagnetic" amplitude fluctuations, respectively.

\begin{figure}[t]
\includegraphics[width=0.8\textwidth]{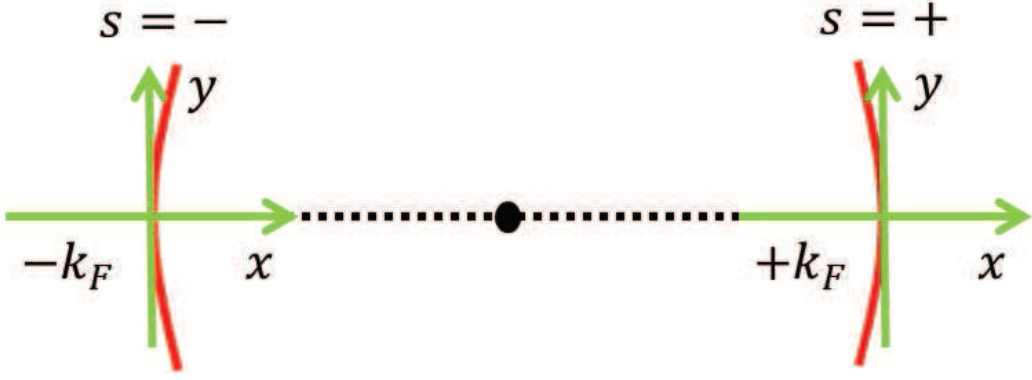}
\caption{A schematic diagram of a Fermi surface in the double patch construction. Red-curved lines denote a pair of Fermi surfaces, connected by $2\bm{k}_{F}$, inside which electrons are filled with. A coordinate system is defined as the figure for each patch of $s = \pm$.} \label{Double_Patch_Construction}
\end{figure}

%
%
%

Resorting to this double patch construction, we obtain an effective theory valid at low energies \bqa && Z = \int D c_{is\sigma} D \delta m_{in} D \bm{n}_{i} \delta(|\bm{n}_{i}|^{2} - 1) \exp\Bigl[ - \int_{0}^{\beta} d \tau \Bigl\{ \sum_{i} c_{is\sigma}^{\dagger} (\partial_{\tau} - \mu) c_{is\sigma} - s t \sum_{ij} (c_{is\sigma}^{\dagger} c_{js\sigma} + H.c.) \nn && - m \sum_{i} c_{is\alpha}^{\dagger} \boldsymbol{n}_{i} \cdot \boldsymbol{\sigma}_{\alpha\beta} c_{i-s\beta} - \sum_{i} \delta m_{i1} c_{is\alpha}^{\dagger} \boldsymbol{n}_{i} \cdot \boldsymbol{\sigma}_{\alpha\beta} c_{is\beta} - \sum_{i} \delta m_{i2} c_{is\alpha}^{\dagger} \boldsymbol{n}_{i} \cdot \boldsymbol{\sigma}_{\alpha\beta} c_{i-s\beta} \nn && + \frac{1}{2g} \sum_{i} \Bigl( \delta m_{in}^{2} + 2 m \delta m_{i2} \Bigr) + L^{d} \frac{1}{2g} m^{2} \Bigr\} \Bigr] . \eqa We emphasize that $c_{is\sigma}$ should be regarded as a low-energy electron field near the Fermi surface, defined on the patch of $s = \pm$. See Fig. 2. We note that ferromagnetic amplitude fluctuations are involved with low-energy electrons in the same patch while antiferromagnetic ones are associated with those in the opposite patch. Although we use the symbol of $\sum_{i}$ with a lattice index $i$, this should be regarded to be just formal. A continuum field theory will be constructed in the next section. We point out that the Fermi velocity has an opposite sign in each patch, denoted by $s ~ t$ in the hopping term, which reflects the $\bm{Q} = 2 \bm{k}_{F}$ antiferromagnetic ordering. If $\bm{Q} \not= 2 \bm{k}_{F}$ is taken into account, the Fermi velocity cannot be parallel to each other.

An idea is to take the strong coupling approach \cite{With_MDKim,FMLQCP}, which diagonalizes the three spin-fermion coupling terms of the second line in Eq. (6). Resorting to the CP$^{1}$ representation \bqa && \boldsymbol{n}_{i} \cdot \boldsymbol{\sigma}_{\alpha\beta} = U_{i\alpha\gamma} \sigma^{3}_{\gamma\delta} U_{i\delta\beta}^{\dagger} , \eqa where \bqa && \bm{U}_{i} = \left(\begin{array}{cc} z_{i\uparrow} & z_{i\downarrow}^{\dagger} \\ z_{i\downarrow} & - z_{i\uparrow}^{\dagger} \end{array} \right) \eqa is an SU(2) matrix field to describe directional spin fluctuations, we introduce a fermion field $f_{is\beta}$, given by the unitary transformation \bqa && c_{is\alpha} = U_{i\alpha\beta} f_{is\beta} . \eqa It is straightforward to rewrite the pre-Hertz-Moriya-Millis theory in terms of the bosonic spinon $z_{i\sigma}$ and the fermionic holon $f_{is\alpha}$ as follows
\bqa && Z = \int D f_{is\sigma} D \delta m_{in} D U_{i\alpha\beta} \delta(\mbox{det} \bm{U}_{i} - 1) \exp\Bigl[ - \int_{0}^{\beta} d \tau \Bigl\{ \sum_{i} f_{is\alpha}^{\dagger} [(\partial_{\tau} - \mu) \delta_{\alpha\beta} - U_{i\alpha\gamma}^{\dagger} \partial_{\tau} U_{i\gamma\beta} ] f_{is\beta} \nn && - s t \sum_{ij} (f_{is\alpha}^{\dagger} U_{i\alpha\gamma}^{\dagger} U_{j\gamma\beta} f_{js\beta} + H.c.) - m \sum_{i} f_{is\alpha}^{\dagger} \sigma^{3}_{\gamma\delta} f_{i-s\beta} - \sum_{i} \delta m_{i1} f_{is\alpha}^{\dagger} \sigma^{3}_{\gamma\delta} f_{is\beta} - \sum_{i} \delta m_{i2} f_{is\alpha}^{\dagger} \sigma^{3}_{\gamma\delta} f_{i-s\beta} \nn && + \frac{1}{2g} \sum_{i} \Bigl( \delta m_{in}^{2} + 2 m \delta m_{i2} \Bigr) + L^{d} \frac{1}{2g} m^{2} \Bigr\} \Bigr] , \eqa where all spin-fermion coupling terms are diagonalized.

The diagonalization procedure gives rise to complex correlations in kinetic-energy terms of spinons and holons. Such correlations between spinons and holons are decomposed, resorting to the Hubbard-Stratonovich transformation \cite{With_MDKim,FMLQCP}. The time part is \bqa && - f_{is\alpha}^{\dagger} U_{i\alpha\gamma}^{\dagger} \partial_{\tau} U_{i\gamma\beta} f_{is\beta} \rightarrow - x_{i} \sigma f_{is\sigma}^{\dagger} f_{is\sigma} + y_{i} z_{i\sigma}^{\dagger} \partial_{\tau} z_{i\sigma} - x_{i} y_{i} , \eqa and the spatial part is \bqa && - s t f_{is\alpha}^{\dagger} U_{i\alpha\gamma}^{\dagger} U_{j\gamma\beta} f_{js\beta} = - s t \Bigl\{ f_{is+}^{\dagger} (z_{i\sigma}^{\dagger} z_{j\sigma}) f_{js+} + f_{is+}^{\dagger} (\epsilon_{\sigma\sigma'} z_{i\sigma}^{\dagger} z_{j\sigma'}^{\dagger}) f_{js-} + f_{is-}^{\dagger} (\epsilon_{\sigma\sigma'} z_{j\sigma} z_{i\sigma'}) f_{js+} + f_{is-}^{\dagger} (z_{j\sigma}^{\dagger} z_{i\sigma}) f_{js-} \Bigr\} \nn && \rightarrow - t \Bigl( s f_{is+}^{\dagger} \chi_{ij}^{f} f_{js+} + z_{i\sigma}^{\dagger} \chi_{ij}^{z +} z_{j\sigma} - \chi_{ij}^{f} \chi_{ij}^{z +} + s f_{is-}^{\dagger} \chi_{ij}^{f *} f_{js-} + z_{j\sigma}^{\dagger} \chi_{ij}^{z -} z_{i\sigma} - \chi_{ij}^{f *} \chi_{ij}^{z -} \Bigr) , \eqa where spin-singlet pairing fluctuations are assumed to be not relevant and neglected. Then, an effective theory reads \bqa && Z = \int D f_{is\sigma} D z_{i\sigma} D \delta m_{in} D x_{i} D y_{i} D \chi_{ij}^{f} D \chi_{ij}^{z \pm} D \lambda_{i} \exp\Bigl[ - \int_{0}^{\beta} d \tau \Bigl\{ \sum_{i} \Bigl( f_{is\sigma}^{\dagger} (\partial_{\tau} - \mu) f_{is\sigma} - \sigma m f_{is\sigma}^{\dagger} f_{i-s\sigma} \nn && - \sigma ( \delta m_{i1} + x_{i}) f_{is\sigma}^{\dagger} f_{is\sigma} - \sigma \delta m_{i2} f_{is\sigma}^{\dagger} f_{i-s\sigma} \Bigr) - s t \sum_{ij} ( f_{is+}^{\dagger} \chi_{ij}^{f} f_{js+} + f_{is-}^{\dagger} \chi_{ij}^{f *} f_{js-}  + H.c. ) + \sum_{i} y_{i} z_{i\sigma}^{\dagger} \partial_{\tau} z_{i\sigma} \nn && - t \sum_{ij} ( z_{i\sigma}^{\dagger} \chi_{ij}^{z +} z_{j\sigma} + z_{j\sigma}^{\dagger} \chi_{ij}^{z -} z_{i\sigma} + H.c. ) + i \sum_{i} \lambda_{i} (|z_{i\sigma}|^{2} - 1) + \frac{1}{2g} \sum_{i} ( \delta m_{in}^{2} + 2 m \delta m_{i2} ) + L^{d} \frac{1}{2g} m^{2} \Bigr\} \nn && + \beta \sum_{i} x_{i} y_{i} - \beta t \sum_{ij} ( \chi_{ij}^{f} \chi_{ij}^{z +} + \chi_{ij}^{f *} \chi_{ij}^{z -} + H.c. ) \Bigr] , \eqa where $\lambda_{i}$ incorporates the unimodular constraint for the spinon field. Although spin-singlet pairing fluctuations are neglected, relevant approximations have not been made. In other words, integration for $x_{i}$, $y_{i}$, $\chi_{ij}^{f}$, and $\chi_{ij}^{z \pm}$ recovers Eq. (10) essentially.

Next, we perform integrals for $x_{i}$ and $y_{i}$. But, we determine hopping parameters of $\chi_{ij}^{f}$ and $\chi_{ij}^{z \pm}$ in the saddle-point approximation, resulting in band renormalization for spinons and holons. The saddle-point value of $i \lambda_{i} \rightarrow \lambda$ plays the role of a mass in spinon excitations. Then, we reach the following expression \bqa && Z = \int D f_{is\sigma} D z_{i\sigma} D \delta m_{in} \exp\Bigl[ - \int_{0}^{\beta} d \tau \Bigl\{ \sum_{i} \Bigl( f_{is\sigma}^{\dagger} (\partial_{\tau} - \mu) f_{is\sigma} - \sigma m f_{is\sigma}^{\dagger} f_{i-s\sigma} - \sigma \delta m_{i1} f_{is\sigma}^{\dagger} f_{is\sigma} - \sigma \delta m_{i2} f_{is\sigma}^{\dagger} f_{i-s\sigma} \Bigr) \nn && - s t \sum_{ij} ( f_{is\sigma}^{\dagger} \chi_{ij}^{f} f_{js\sigma} + H.c ) + \frac{1}{2g} \sum_{i} ( z_{i\sigma}^{\dagger} \partial_{\tau} z_{i\sigma} - \delta m_{i1} )^{2} - t \sum_{ij} ( z_{i\sigma}^{\dagger} \chi_{ij}^{z} z_{j\sigma} + H.c. ) + \lambda \sum_{i} |z_{i\sigma}|^{2} \nn && + \frac{1}{2g} \sum_{i} ( \delta m_{i2}^{2} + 2 m \delta m_{i2} ) + L^{d} \frac{1}{2g} m^{2} \Bigr\} - \beta \Bigl\{ t \sum_{ij} ( \chi_{ij}^{f} \chi_{ij}^{z} + H.c. ) - L^{d} \lambda \Bigr\} \Bigr] , \eqa where the saddle-point analysis for hopping parameters of spinons gives \bqa && \chi_{ij}^{z +} = \chi_{ij}^{z - *} = \chi_{ij}^{z} . \eqa

We call this formulation U(1) slave spin-rotor theory, the name of which is to benchmark U(1) slave-rotor theory for charge fluctuations \cite{U1SCR}. Unfortunately, U(1) slave spin-rotor theory turns out to be not stable in contrast with U(1) slave charge-rotor theory. The positive sign in $\frac{1}{2g} \sum_{i} ( z_{i\sigma}^{\dagger} \partial_{\tau} z_{i\sigma} - \delta m_{i1} )^{2}$ favors stronger directional fluctuations while it is negative in the U(1) slave-rotor theory, serving a parabolic potential for charge fluctuations. This difference originates from the opposite sign when the Hubbard-$U$ term is decomposed into charge and spin channels.

\subsection{Integration of amplitude fluctuations}

An idea overcoming the inconsistency in dynamics of transverse spin fluctuations is to introduce quantum corrections from amplitude (ferromagnetic) fluctuations, renormalizing their dynamics. Taking the Luttinger-Ward functional approach \cite{LW_Functional} within the Eliashberg approximation \cite{Kim_Pepin_LW}, we construct the partition function as follows
\bqa && Z = \int D f_{is\sigma} D z_{i\sigma} \exp\Bigl[ - \int_{0}^{\beta} d \tau \Bigl\{ \sum_{i} \Bigl( f_{is\sigma}^{\dagger} (\partial_{\tau} - \mu) f_{is\sigma} - \sigma m f_{is\sigma}^{\dagger} f_{i-s\sigma} \Bigr) - s t \sum_{ij} ( f_{is\sigma}^{\dagger} \chi_{ij}^{f} f_{js\sigma} + H.c ) \nn && - \frac{1}{2} \int_{0}^{\beta} d \tau' \sum_{i} \sum_{i'} \Bigl(\sigma f_{is\sigma}^{\dagger} f_{is\sigma} + \frac{1}{g} z_{i\sigma}^{\dagger} \partial_{\tau} z_{i\sigma} \Bigr)_{\tau} D_{ii'}^{(1)} (\tau-\tau';m) \Bigl(\sigma' f_{i's'\sigma'}^{\dagger} f_{i's'\sigma'} + \frac{1}{g} z_{i'\sigma'}^{\dagger} \partial_{\tau'} z_{i'\sigma'} \Bigr)_{\tau'} \nn && - \frac{1}{2} \int_{0}^{\beta} d \tau' \sum_{i} \sum_{i'} \Bigl(\sigma f_{is\sigma}^{\dagger} f_{i-s\sigma} - \frac{1}{g} m \Bigr)_{\tau} D_{ii'}^{(2)} (\tau-\tau';m) \Bigl(\sigma' f_{i's'\sigma'}^{\dagger} f_{i'-s'\sigma'} - \frac{1}{g} m \Bigr)_{\tau'} \Bigr\} \nn && - \frac{1}{2} \sum_{i\Omega} \sum_{\boldsymbol{q}} \Bigl\{ \ln \Bigl( \frac{1}{4g} - \Pi^{(1)}(\boldsymbol{q},i\Omega;m) \Bigr) + \Pi^{(1)}(\boldsymbol{q},i\Omega;m) D^{(1)}(\boldsymbol{q},i\Omega;m) \Bigr\} \nn && - \frac{1}{2} \sum_{i\Omega} \sum_{\boldsymbol{q}} \Bigl\{ \ln \Bigl( \frac{1}{4g} - \Pi^{(2)}(\boldsymbol{q},i\Omega;m) \Bigr) + \Pi^{(2)}(\boldsymbol{q},i\Omega;m) D^{(2)}(\boldsymbol{q},i\Omega;m) \Bigr\} \nn && - \int_{0}^{\beta} d \tau \Bigl\{ \frac{1}{2g} \sum_{i} ( z_{i\sigma}^{\dagger} \partial_{\tau} z_{i\sigma} )^{2} - t \sum_{ij} ( z_{i\sigma}^{\dagger} \chi_{ij}^{z} z_{j\sigma} + H.c. ) + \lambda \sum_{i} |z_{i\sigma}|^{2} \Bigr\} \nn && - \beta \Bigl\{ t \sum_{ij} ( \chi_{ij}^{f} \chi_{ij}^{z} + H.c. ) + L^{d} \frac{1}{2g} m^{2} - L^{d} \lambda \Bigr\} \Bigr] , \eqa
where $D^{(n)}(\boldsymbol{q},i\Omega;m) = \frac{1}{\frac{1}{4g} - \Pi^{(n)}(\boldsymbol{q},i\Omega;m)}$ with $n = 1, 2$ is the propagator of ferromagnetic and antiferromagnetic amplitude fluctuations, respectively, and $\Pi^{(n)}(\boldsymbol{q},i\Omega;m)$ is the self-energy of the amplitude-fluctuation propagator, given by the fermion polarization bubble.

Performing the Fourier transformation, we obtain \bqa && Z = \int D f_{\bm{k}s\sigma} D z_{\bm{k}\sigma} \exp\Bigl[ - \Bigl\{ \sum_{i\omega} \sum_{\bm{k}} \Bigl( f_{\bm{k}s\sigma}^{\dagger} (- i \omega - \mu - z s t \chi^{f} \gamma_{\boldsymbol{k}}) f_{\bm{k}s\sigma} + \sigma m \frac{\Pi^{(2)}(m)}{\frac{1}{4g} - \Pi^{(2)}(m)} f_{\bm{k}s\sigma}^{\dagger} f_{\bm{k}-s\sigma} \Bigr) \nn && - \frac{1}{2} \sum_{i\Omega} \frac{1}{\beta} \sum_{i\omega} \frac{1}{\beta} \sum_{i\omega'} \sum_{\boldsymbol{q}} \sum_{\boldsymbol{k}} \sum_{\boldsymbol{k}'} [\sigma f_{\boldsymbol{k}s\sigma}^{\dagger}(i\omega) f_{\boldsymbol{k}+\boldsymbol{q}s\sigma}(i\omega+i\Omega)] D^{(1)}(\boldsymbol{q},i\Omega;m) [\sigma' f_{\boldsymbol{k}'s'\sigma'}^{\dagger}(i\omega') f_{\boldsymbol{k}'-\boldsymbol{q}s'\sigma'}(i\omega'-i\Omega)] \nn && - \frac{1}{2} \sum_{i\Omega} \frac{1}{\beta} \sum_{i\omega} \frac{1}{\beta} \sum_{i\omega'} \sum_{\boldsymbol{q}} \sum_{\boldsymbol{k}} \sum_{\boldsymbol{k}'} [\sigma f_{\boldsymbol{k}s\sigma}^{\dagger}(i\omega) f_{\boldsymbol{k}+\boldsymbol{q}-s\sigma}(i\omega+i\Omega)] D^{(2)}(\boldsymbol{q},i\Omega;m) [\sigma' f_{\boldsymbol{k}'s'\sigma'}^{\dagger}(i\omega') f_{\boldsymbol{k}'-\boldsymbol{q}-s'\sigma'}(i\omega'-i\Omega)] \nn && + \frac{1}{g} \sum_{i\Omega} \frac{1}{\beta} \sum_{i\omega} \frac{1}{\beta} \sum_{i\omega'} \sum_{\boldsymbol{q}} \sum_{\boldsymbol{k}} \sum_{\boldsymbol{k}'} [\sigma f_{\boldsymbol{k}s\sigma}^{\dagger}(i\omega) f_{\boldsymbol{k}+\boldsymbol{q}s\sigma}(i\omega+i\Omega)] D^{(1)}(\boldsymbol{q},i\Omega;m) \Bigl(i\omega'-\frac{1}{2}i\Omega\Bigr) [ z_{\boldsymbol{k}'\sigma'}^{\dagger}(i\omega') z_{\boldsymbol{k}'-\boldsymbol{q}\sigma'}(i\omega'-i\Omega) ] \Bigr\} \nn && - \frac{1}{2} \sum_{i\Omega} \sum_{\boldsymbol{q}} \Bigl\{ \ln \Bigl( \frac{1}{4g} - \Pi^{(1)}(\boldsymbol{q},i\Omega;m) \Bigr) + \Pi^{(1)}(\boldsymbol{q},i\Omega;m) D^{(1)}(\boldsymbol{q},i\Omega;m) \Bigr\} \nn && - \frac{1}{2} \sum_{i\Omega} \sum_{\boldsymbol{q}} \Bigl\{ \ln \Bigl( \frac{1}{4g} - \Pi^{(2)}(\boldsymbol{q},i\Omega;m) \Bigr) + \Pi^{(2)}(\boldsymbol{q},i\Omega;m) D^{(2)}(\boldsymbol{q},i\Omega;m) \Bigr\} \nn && - \sum_{i\Omega} \sum_{\boldsymbol{q}} \Bigl\{ - \frac{1}{4g} ( z_{i\sigma}^{\dagger} \partial_{\tau} z_{i\sigma} )_{\boldsymbol{q},i\Omega} \frac{\Pi^{(1)}(\boldsymbol{q},i\Omega;m)}{\frac{1}{4g} - \Pi^{(1)}(\boldsymbol{q},i\Omega;m)} ( z_{i'\sigma'}^{\dagger} \partial_{\tau'} z_{i'\sigma'} )_{-\boldsymbol{q},-i\Omega} + ( \lambda - z t \chi^{z} \gamma_{\boldsymbol{q}} ) z_{\boldsymbol{q}\sigma}^{\dagger} z_{\boldsymbol{q}\sigma} \Bigr\} \nn && - \beta L^{d} \Bigl( 2 z t \chi^{f} \chi^{z} - \lambda - \frac{m^{2}}{8g} \frac{\Pi^{(2)}(m)}{\frac{1}{4g} - \Pi^{(2)}(m)} \Bigr) \Bigr] , \eqa where $z t \chi^{f(z)} \gamma_{\bm{k}}$ is a Fourier-transformed expression of $\chi_{ij}^{f(z)} t_{ij}$ with a coordination number $z$, taking $\chi_{ij}^{f} = \chi^{f}$ and $\chi_{ij}^{z} = \chi^{z}$ in the saddle-point approximation. First of all, the inconsistency for dynamics of transverse spin fluctuations is resolved by the renormalization of ferromagnetic amplitude fluctuations. In the temporal part of transverse spin fluctuations, $- \frac{1}{4g} ( z_{i\sigma}^{\dagger} \partial_{\tau} z_{i\sigma} )_{\boldsymbol{q},i\Omega} \frac{\Pi^{(1)}(\boldsymbol{q},i\Omega;m)}{\frac{1}{4g} - \Pi^{(1)}(\boldsymbol{q},i\Omega;m)} ( z_{i'\sigma'}^{\dagger} \partial_{\tau'} z_{i'\sigma'} )_{-\boldsymbol{q},-i\Omega}$, we observe the sign change from $\frac{1}{2g} > 0$ to $- \frac{1}{4g} \frac{\Pi^{(1)}(\boldsymbol{q},i\Omega;m)}{\frac{1}{4g} - \Pi^{(1)}(\boldsymbol{q},i\Omega;m)} < 0$, where $\frac{1}{4g} - \Pi^{(1)}(\boldsymbol{q},i\Omega;m) > 0$. As a result, dynamics of transverse spin fluctuations is well defined as it must be. It is also noticeable that the magnetization order parameter is renormalized to $m \frac{\Pi^{(2)}(m)}{\frac{1}{4g} - \Pi^{(2)}(m)}$.

We would like to compare this effective theory with the Hertz-Moriya-Millis theory, given by \bqa && Z = \int D c_{\bm{k}s\sigma} \exp\Bigl[ - \Bigl\{ \sum_{i\omega} \sum_{\bm{k}} \Bigl( c_{\bm{k}s\sigma}^{\dagger} (- i \omega - \mu - z s t \gamma_{\boldsymbol{k}}) c_{\bm{k}s\sigma} + \sigma m \frac{\Pi^{(2)}(m)}{\frac{1}{4g} - \Pi^{(2)}(m)} c_{\bm{k}s\sigma}^{\dagger} c_{\bm{k}-s\sigma} \Bigr) \nn && - \frac{1}{2} \sum_{i\Omega} \frac{1}{\beta} \sum_{i\omega} \frac{1}{\beta} \sum_{i\omega'} \sum_{\boldsymbol{q}} \sum_{\boldsymbol{k}} \sum_{\boldsymbol{k}'} [\sigma c_{\boldsymbol{k}s\sigma}^{\dagger}(i\omega) c_{\boldsymbol{k}+\boldsymbol{q}s\sigma}(i\omega+i\Omega)] D^{(1)}(\boldsymbol{q},i\Omega;m) [\sigma' c_{\boldsymbol{k}'s'\sigma'}^{\dagger}(i\omega') c_{\boldsymbol{k}'-\boldsymbol{q}s'\sigma'}(i\omega'-i\Omega)] \nn && - \frac{1}{2} \sum_{i\Omega} \frac{1}{\beta} \sum_{i\omega} \frac{1}{\beta} \sum_{i\omega'} \sum_{\boldsymbol{q}} \sum_{\boldsymbol{k}} \sum_{\boldsymbol{k}'} [\sigma c_{\boldsymbol{k}s\sigma}^{\dagger}(i\omega) c_{\boldsymbol{k}+\boldsymbol{q}-s\sigma}(i\omega+i\Omega)] D^{(2)}(\boldsymbol{q},i\Omega;m) [\sigma' c_{\boldsymbol{k}'s'\sigma'}^{\dagger}(i\omega') c_{\boldsymbol{k}'-\boldsymbol{q}-s'\sigma'}(i\omega'-i\Omega)] \Bigr\} \nn && - \frac{1}{2} \sum_{i\Omega} \sum_{\boldsymbol{q}} \Bigl\{ \ln \Bigl( \frac{1}{4g} - \Pi^{(1)}(\boldsymbol{q},i\Omega;m) \Bigr) + \Pi^{(1)}(\boldsymbol{q},i\Omega;m) D^{(1)}(\boldsymbol{q},i\Omega;m) \Bigr\} \nn && - \frac{1}{2} \sum_{i\Omega} \sum_{\boldsymbol{q}} \Bigl\{ \ln \Bigl( \frac{1}{4g} - \Pi^{(2)}(\boldsymbol{q},i\Omega;m) \Bigr) + \Pi^{(2)}(\boldsymbol{q},i\Omega;m) D^{(2)}(\boldsymbol{q},i\Omega;m) \Bigr\} + \beta L^{d} \frac{m^{2}}{8g} \frac{\Pi^{(2)}(m)}{\frac{1}{4g} - \Pi^{(2)}(m)} \Bigr] , \nonumber \eqa where only amplitude fluctuations are taken into account. In this respect U(1) slave spin-rotor theory deals with not only amplitude fluctuations but also transverse spin fluctuations, consistently. In this study we show that an effective interaction between renormalized electrons and transverse spin fluctuations, that is, $\frac{1}{g} \sum_{i\Omega} \frac{1}{\beta} \sum_{i\omega} \frac{1}{\beta} \sum_{i\omega'} \sum_{\boldsymbol{q}} \sum_{\boldsymbol{k}} \sum_{\boldsymbol{k}'} [\sigma f_{\boldsymbol{k}s\sigma}^{\dagger}(i\omega) f_{\boldsymbol{k}+\boldsymbol{q}s\sigma}(i\omega+i\Omega)] D^{(1)}(\boldsymbol{q},i\Omega;m) \Bigl(i\omega'-\frac{1}{2}i\Omega\Bigr) [ z_{\boldsymbol{k}'\sigma'}^{\dagger}(i\omega') z_{\boldsymbol{k}'-\boldsymbol{q}\sigma'}(i\omega'-i\Omega) ]$, gives rise to local quantum criticality in an antiferromagnetic quantum phase transition, characterized by the emergence of localized magnetic moments.

\begin{figure}[t]
\includegraphics[width=0.8\textwidth]{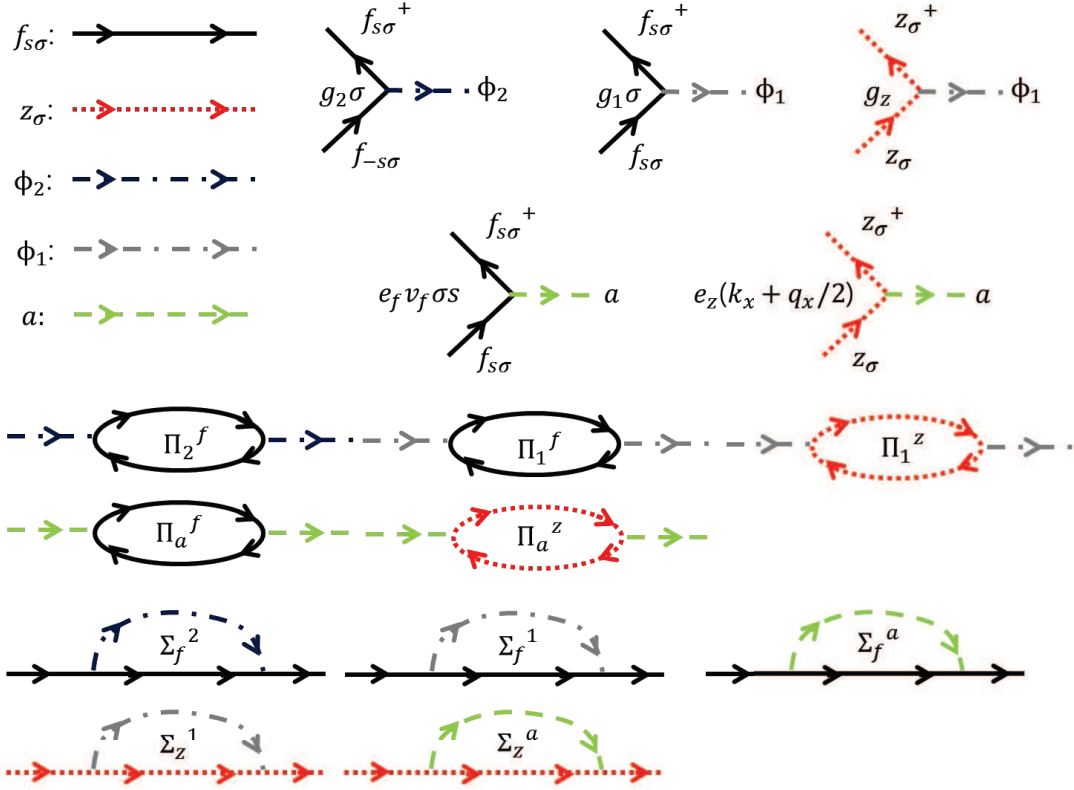}
\caption{Feynmann diagrams in the Eliashberg approximation. Renormalized electrons $f_{s\sigma}$ and longitudinal antiferromagnetic fluctuations $\phi_{2}$ consist of the Hertz-Moriya-Millis theory while transverse spin fluctuations $z_{\sigma}$ and U(1) gauge fields $a$ appear in the U(1) slave spin-rotor formulation. Longitudinal ferromagnetic fluctuations $\phi_{1}$ give rise to consistency for dynamics of transverse spin fluctuations $z_{\sigma}$. All interaction vertices are given by diagrams with three lines, which characterize the U(1) slave spin-rotor theory. Self-energy corrections of three bosonic fields, $\phi_{2}$, $\phi_{1}$, and $a$ are given by polarization bubbles of renormalized electrons and transverse spin fluctuations. Self-energy corrections of renormalized electrons and transverse spin fluctuations are given by typical one-loop diagrams with a bosonic line. Since longitudinal ferromagnetic fluctuations are gapped, one may neglect all diagrams involved with $\phi_{1}$ except for the spinon self-energy, where it makes dynamics of spinons consistent. In the lattice-model construction, dynamics of gauge fluctuations are neglected, not incorporated into the free energy.} \label{Feynmann_Diagrams}
\end{figure}

It is straightforward to obtain the Luttinger-Ward free-energy functional in the Eliashberg approximation \cite{Kim_Pepin_LW} \bqa && F(m,\lambda;\mu,g,T) \nn && = - \frac{N_{\sigma}}{\beta} \sum_{i\omega} \sum_{\bm{k}} \mbox{tr} \Bigl[ \ln \Bigl\{ \left( \begin{array} [c]{cc} - i \omega - \mu - z t \chi^{f} \gamma_{\boldsymbol{k}} & \sigma m \frac{\Pi^{(2)}(m)}{\frac{1}{4g} - \Pi^{(2)}(m)} \\ \sigma m \frac{\Pi^{(2)}(m)}{\frac{1}{4g} - \Pi^{(2)}(m)} & - i \omega - \mu + z t \chi_{f} \gamma_{\boldsymbol{k}} \end{array} \right) + \boldsymbol{\Sigma}_{f}(\boldsymbol{k},i\omega;m) \Bigr\} + \boldsymbol{\Sigma}_{f}(\boldsymbol{k},i\omega;m) \boldsymbol{G}_{f}(\boldsymbol{k},i\omega;m) \Bigr] \nn && + \frac{N_{\sigma}}{\beta} \sum_{i\omega} \sum_{\bm{k}} \ln \Bigl\{ \lambda - z t \chi^{z} \gamma_{\bm{k}} + \Sigma_{z}(\boldsymbol{k},i\omega;m) \Bigr\} + \frac{1}{2\beta} \sum_{i\Omega} \sum_{\boldsymbol{q}} \ln \Bigl( \frac{1}{4g} - \Pi^{(1)}(\boldsymbol{q},i\Omega;m) \Bigr) + \frac{1}{2\beta} \sum_{i\Omega} \sum_{\boldsymbol{q}} \ln \Bigl( \frac{1}{4g} - \Pi^{(2)}(\boldsymbol{q},i\Omega;m) \Bigr) \nn && + L^{d} \Bigl\{ 2 z t \chi^{f} \chi^{z} - \lambda - \frac{m^{2}}{8g} \frac{\Pi^{(2)}(m)}{\frac{1}{4g} - \Pi^{(2)}(m)} \Bigr\} , \eqa where \bqa && \Pi^{(1)}(\boldsymbol{q},i\Omega;m) = \frac{N_{\sigma}}{\beta} \sum_{i\omega} \sum_{\boldsymbol{k}} \Bigl\{ \boldsymbol{G}_{f}^{11}(\boldsymbol{k} + \boldsymbol{q},i\omega + i \Omega;m) \boldsymbol{G}_{f}^{11}(\boldsymbol{k},i\omega;m) + \boldsymbol{G}_{f}^{22}(\boldsymbol{k} + \boldsymbol{q},i\omega + i \Omega;m) \boldsymbol{G}_{f}^{22}(\boldsymbol{k},i\omega;m) \Bigr\} , \nn && \Pi^{(2)}(\boldsymbol{q},i\Omega;m) = \frac{N_{\sigma}}{\beta} \sum_{i\omega} \sum_{\boldsymbol{k}} \Bigl\{ \boldsymbol{G}_{f}^{11}(\boldsymbol{k} + \boldsymbol{q},i\omega + i \Omega;m) \boldsymbol{G}_{f}^{22}(\boldsymbol{k},i\omega;m) + \boldsymbol{G}_{f}^{22}(\boldsymbol{k} + \boldsymbol{q},i\omega + i \Omega;m) \boldsymbol{G}_{f}^{11}(\boldsymbol{k},i\omega;m) \Bigr\} , \nn && \boldsymbol{\Sigma}_{f}(\boldsymbol{k},i\omega;m) = - \frac{1}{\beta} \sum_{i\Omega} \sum_{\boldsymbol{q}} \boldsymbol{G}_{f}(\boldsymbol{k} + \boldsymbol{q},i\omega + i \Omega;m) D^{(1)}(\boldsymbol{q},i\Omega;m) - \frac{1}{\beta} \sum_{i\Omega} \sum_{\boldsymbol{q}} \boldsymbol{G}_{f}(\boldsymbol{k} + \boldsymbol{q},i\omega + i \Omega;m) D^{(2)}(\boldsymbol{q},i\Omega;m) , \nn && \Sigma_{z}(\boldsymbol{k},i\omega;m) = - \frac{1}{\beta} \sum_{i\Omega} \sum_{\bm{q}} G_{z}(\bm{k}-\bm{q},i\omega-i\Omega;m) \frac{(i\omega+i\Omega/2)^{2} \Pi^{(1)}(\boldsymbol{q},i\Omega;m)}{1 - 4 g \Pi^{(1)}(\boldsymbol{q},i\Omega;m)} \eqa are self-energies of ferromagnetic, antiferromagnetic amplitude fluctuations, renormalized electrons, and transverse spin fluctuations, respectively, and \bqa && \boldsymbol{G}_{f}(\boldsymbol{k},i\omega;m) = - \Bigl\{ \left( \begin{array} [c]{cc} - i \omega - \mu - z t \chi^{f} \gamma_{\boldsymbol{k}} & \sigma m \frac{\Pi^{(2)}(m)}{\frac{1}{4g} - \Pi^{(2)}(m)} \\ \sigma m \frac{\Pi^{(2)}(m)}{\frac{1}{4g} - \Pi^{(2)}(m)} & - i \omega - \mu + z t \chi^{f} \gamma_{\boldsymbol{k}} \end{array} \right) + \boldsymbol{\Sigma}_{f}(\boldsymbol{k},i\omega;m) \Bigr\}^{-1} , \nn && D^{(1)}(\boldsymbol{q},i\Omega;m) = \frac{1}{\frac{1}{4g} - \Pi^{(1)}(\boldsymbol{q},i\Omega;m)} , ~~~~~ D^{(2)}(\boldsymbol{q},i\Omega;m) = \frac{1}{\frac{1}{4g} - \Pi^{(2)}(\boldsymbol{q},i\Omega;m)} , \nn && G_{z}(\bm{k},i\omega;m) = \frac{1}{\lambda - z t \chi^{z} \gamma_{\bm{k}} + \Sigma_{z}(\boldsymbol{k},i\omega;m)} \eqa are Green's functions of holons, ferromagnetic, antiferromagnetic amplitude fluctuations, and spinons, respectively. The self-energy of ferromagnetic (antiferromagnetic) amplitude fluctuations is given by the polarization bubble of renormalized electrons within the same (opposite) patch. The holon self-energy results from scattering with both amplitude fluctuations, and the spinon self-energy originates from the renormalization by ferromagnetic amplitude fluctuations. All quantum corrections in the Eliashberg approximation are shown in Fig. 3.

Performing the mean-field analysis for the magnetization order parameter $m$ and the mass parameter of spinons $\lambda$, one can find a phase diagram for an antiferromagnetic quantum phase transition, where the condensation transition of spinons is expected to allow novel physics. Generally speaking, one may speculate that the antiferromagnetic transition given by the formation of an antiferromagnetic order parameter $m$ will not coincide with the condensation transition of spinons given by $\lambda = 0$. Although this scenario may be possible, we suggests another scenario based on the renormalization group analysis of an effective field theory within the U(1) slave spin-rotor representation, where both transitions meet at one point. The spinon-holon interaction vertex, not taken into account in the Luttinger-Ward free-energy functional, turns out to play a central role in the emergence of
localized magnetic moments at an antiferromagnetic quantum critical point.

\section{Renormalization group analysis}

\subsection{A field theory for antiferromagnetic quantum criticality approaching from the Fermi-liquid state}

Following the patch construction of Ref. \cite{SSL_U1GT}, we write down an effective field theory in the U(1) slave spin-rotor representation \bqa && Z = \int D f_{s\sigma} D z_{\sigma} D \phi_{n} D a \exp\Bigl[ - \int_{0}^{\beta} d \tau \int_{-\infty}^{\infty} d x \int_{-\infty}^{\infty} d y \Bigl\{ f_{s\sigma}^{\dagger} \Bigl(\partial_{\tau} - i s v_{F} \partial_{x} - \frac{v_{F}}{2\gamma} \partial_{y}^{2}\Bigr) f_{s\sigma} \nn && + \phi_{1} \Bigl( \partial_{\tau} - v_{1}^{2} \partial_{x}^{2} - v_{1}^{2} \partial_{y}^{2} + m_{1}^{2} \Bigr) \phi_{1} + \frac{u_{1}}{2} \phi_{1}^{4} + \phi_{2} \Bigl( - \partial_{\tau}^{2} - v_{2}^{2} \partial_{x}^{2} - v_{2}^{2} \partial_{y}^{2} + m_{2}^{2} \Bigr) \phi_{2} + \frac{u_{2}}{2} \phi_{2}^{4} + a ( - \partial_{\tau}^{2} - v_{a}^{2} \partial_{x}^{2} - v_{a}^{2} \partial_{y}^{2} ) a \nn && - g_{1} \phi_{1} \sigma f_{s\sigma}^{\dagger} f_{s\sigma} - g_{2} \phi_{2} \sigma f_{s\sigma}^{\dagger} f_{-s\sigma}  - e_{f} s v_{F} a \sigma f_{s\sigma}^{\dagger} f_{s\sigma} - g_{z} \phi_{1} z_{\sigma}^{\dagger} \partial_{\tau} z_{\sigma} + z_{\sigma}^{\dagger} ( - v_{z}^{2} \partial_{x}^{2} - v_{z}^{2} \partial_{y}^{2} + m_{z}^{2} ) z_{\sigma} + \frac{u_{z}}{2} |z_{\sigma}|^{4} \nn && - i e_{z} a [ z_{\sigma}^{\dagger} (\partial_{x} z_{\sigma}) - (\partial_{x} z_{\sigma}^{\dagger}) z_{\sigma}] \Bigr\} \Bigr] , \eqa regarded to be a continuum version of Eq. (14). $f_{s\sigma}$ is a low-energy renormalized electron field (holon) with spin $\sigma$ on the Fermi surface of a $s = \pm$ patch. Its dispersion relation is given by $\epsilon(k_{\parallel},k_{\perp}) = s v_{F} k_{\parallel} + \frac{v_{F}}{2\gamma} k_{\perp}^{2}$, where $k_{\parallel}$ is the longitudinal momentum out of the Fermi surface and $k_{\perp}$ is the transverse momentum along the Fermi surface. $v_{F}$ is a Fermi velocity and $\gamma$ is a Landau-damping coefficient \cite{U1GT_Scaling}. See Fig. 2 for our coordinate system. $\phi_{n}$ with $n = 1, 2$ represent ferromagnetic and antiferromagnetic amplitude fluctuations, where their dispersion relations are given by non-relativistic $E_{1}(k_{\parallel},k_{\perp}) = v_{1}^{2} (k_{\parallel}^{2} + k_{\perp}^{2}) + m_{1}^{2}$ and relatisvistic $E_{2}(k_{\parallel},k_{\perp}) = \pm \sqrt{v_{2}^{2} (k_{\parallel}^{2} + k_{\perp}^{2}) + m_{2}^{2}}$, respectively, although these bare dispersions are not much relevant for their renormalized dynamics. Since only antiferromagnetic amplitude fluctuations are allowed to be critical, we assume safely that ferromagnetic amplitude fluctuations are gapped ($m_{1} \not= 0$) but antiferromagnetic ones are gapless ($m_{2} = 0$) at the antiferromagnetic quantum critical point. $u_{n}$ with $n = 1, 2$ are their self-interaction constants. $a$ is a transverse gauge field with the relativistic dispersion $E_{a}(k_{\parallel},k_{\perp}) = v_{a} \sqrt{ k_{\parallel}^{2} + k_{\perp}^{2} }$. $z_{\sigma}$ is a transverse spin-fluctuation field (spinon), where the temporal part is given by the one-loop correction from gapped ferromagnetic amplitude fluctuations, giving rise to consistency for their dynamics as discussed in the last section. $v_{z}$ and $m_{z}$ are the velocity and mass of spinons, respectively. An essential feature of the present study is that both the velocity and mass of spinons become renormalized to vanish at the antiferromagnetic quantum critical point of $m_{2} = 0$, identified with local quantum criticality. $u_{z}$ is the self-interaction parameter of spinons. Gapped ferromagnetic amplitude fluctuations couple to both holons and spinons with coupling constants of $g_{1}$ and $g_{z}$, respectively. Critical antiferromagnetic amplitude fluctuations couple to only holons with $g_{2}$ while gapless gauge fluctuations do to both holons and spinons with $e_{f}$ and $e_{z}$, respectively.

Next, we introduce quantum corrections into this field theory within the Eliashberg approximation as discussed in the last section. All quantum corrections in the one-loop level are shown in Fig. 3. Then, we obtain \bqa && Z = \int D f_{s\sigma} D z_{\sigma} D \phi_{n} D a \exp\Bigl[ - \int_{-\infty}^{\infty} \frac{d \omega}{2\pi} \int_{-\infty}^{\infty} \frac{d k_{\parallel}}{2\pi} \int_{-\infty}^{\infty} \frac{d k_{\perp}}{2\pi} \Bigl\{ f_{s\sigma}^{\dagger} \Bigl(- i \omega - s v_{F} k_{\parallel} - \frac{v_{F}}{2\gamma} k_{\perp}^{2} - i \frac{c_{2} \mbox{sgn}(\omega)}{N_{\sigma}} |\omega|^{\frac{1}{2}} \nn && - i \frac{c_{a} \mbox{sgn}(\omega)}{N_{\sigma}} |\omega|^{\frac{2}{3}} \Bigr) f_{s\sigma} + \phi_{1} \Bigl( \gamma \frac{|\omega|}{|k_{\perp}|} - i \omega + v_{1}^{2} k_{\parallel}^{2} + v_{1}^{2} k_{\perp}^{2} + m_{1}^{2} \Bigr) \phi_{1} + \phi_{2} \Bigl( \gamma_{2} |\omega| + \omega^{2} + v_{2}^{2} k_{\parallel}^{2} + v_{2}^{2} k_{\perp}^{2} \Bigr) \phi_{2} \nn && + a \Bigl( \gamma \frac{|\omega|}{|k_{\perp}|} + \omega^{2} + v_{a}^{2} k_{\parallel}^{2} + v_{a}^{2} k_{\perp}^{2} \Bigr) a - \frac{g_{z}^{2}}{2N_{\sigma}} \int_{-\infty}^{\infty} \frac{d \omega'}{2\pi} \int_{-\infty}^{\infty} \frac{d k_{\parallel}'}{2\pi} \int_{-\infty}^{\infty} \frac{d k_{\perp}'}{2\pi} \int_{-\infty}^{\infty} \frac{d \Omega}{2\pi} \int_{-\infty}^{\infty} \frac{d q_{\parallel}}{2\pi} \int_{-\infty}^{\infty} \frac{d q_{\perp}}{2\pi} \nn && z_{\sigma}^{\dagger}(\omega+\Omega,k_{\parallel}+q_{\parallel},k_{\perp}+q_{\perp}) z_{\sigma}(\omega,k_{\parallel},k_{\perp}) \frac{\Bigl(i \omega + i \frac{\Omega}{2}\Bigr) \Bigl(i \omega' - i \frac{\Omega}{2}\Bigr) }{\gamma \frac{|\Omega|}{|q_{\perp}|} - i \Omega + v_{1}^{2} q_{\parallel}^{2} + v_{1}^{2} q_{\perp}^{2} + m_{1}^{2}} z_{\sigma'}^{\dagger}(\omega'-\Omega,k_{\parallel}'-q_{\parallel},k_{\perp}'-q_{\perp}) z_{\sigma'}(\omega',k_{\parallel}',k_{\perp}') \nn && + z_{\sigma}^{\dagger} ( v_{z}^{2} k_{\parallel}^{2} + v_{z}^{2} k_{\perp}^{2} + m_{z}^{2} ) z_{\sigma} - \frac{g_{1} g_{z}}{N_{\sigma}} \int_{-\infty}^{\infty} \frac{d \omega'}{2\pi} \int_{-\infty}^{\infty} \frac{d k_{\parallel}'}{2\pi} \int_{-\infty}^{\infty}\frac{d k_{\perp}}{2\pi} \int_{-\infty}^{\infty} \frac{d \Omega}{2\pi} \int_{-\infty}^{\infty} \frac{d q_{\parallel}}{2\pi} \int_{-\infty}^{\infty} \frac{d q_{\perp}}{2\pi} \nn && \sigma f_{s\sigma}^{\dagger}(\omega+\Omega,k_{\parallel}+q_{\parallel},k_{\perp}+q_{\perp}) f_{s\sigma}(\omega,k_{\parallel},k_{\perp}) \frac{ i \omega' - i \frac{\Omega}{2} }{\gamma \frac{|\Omega|}{|q_{\perp}|} - i \Omega + v_{1}^{2} q_{\parallel}^{2} + v_{1}^{2} q_{\perp}^{2} + m_{1}^{2}} z_{\sigma'}^{\dagger}(\omega'-\Omega,k_{\parallel}'-q_{\parallel},k_{\perp}'-q_{\perp}) z_{\sigma'}(\omega',k_{\parallel}',k_{\perp}') \nn && - \int_{-\infty}^{\infty} \frac{d \Omega}{2\pi} \int_{-\infty}^{\infty} \frac{d q_{\parallel}}{2\pi} \int \frac{d q_{\perp}}{2\pi} \Bigl( \frac{e_{f}}{\sqrt{N_{\sigma}}} s a(\Omega,q_{\parallel},q_{\perp}) v_{F} \sigma f_{s\sigma}^{\dagger}(\omega+\Omega,k_{\parallel}+q_{\parallel},k_{\perp}+q_{\perp}) f_{s\sigma}(\omega,k_{\parallel},k_{\perp}) \nn && + \frac{g_{1}}{\sqrt{N_{\sigma}}} \phi_{1}(\Omega,q_{\parallel},q_{\perp}) \sigma f_{s\sigma}^{\dagger}(\omega+\Omega,k_{\parallel}+q_{\parallel},k_{\perp}+q_{\perp}) f_{s\sigma}(\omega,k_{\parallel},k_{\perp}) \nn && + \frac{g_{2}}{\sqrt{N_{\sigma}}} \phi_{2}(\Omega,q_{\parallel},q_{\perp}) \sigma f_{s\sigma}^{\dagger}(\omega+\Omega,k_{\parallel}+q_{\parallel},k_{\perp}+q_{\perp}) f_{-s\sigma}(\omega,k_{\parallel},k_{\perp}) \nn && + \frac{e_{z}}{\sqrt{N_{\sigma}}} a(\Omega,q_{\parallel},q_{\perp}) \Bigl( k_{\parallel} + \frac{q_{\parallel}}{2} \Bigr) z_{\sigma}^{\dagger}(\omega+\Omega,k_{\parallel}+q_{\parallel},k_{\perp}+q_{\perp}) z_{\sigma}(\omega,k_{\parallel},k_{\perp}) \nn && + \frac{g_{z}}{\sqrt{N_{\sigma}}} \phi_{1}(\Omega,q_{\parallel},q_{\perp}) \Bigl( i \omega + i \frac{\Omega}{2} \Bigr) z_{\sigma}^{\dagger}(\omega+\Omega,k_{\parallel}+q_{\parallel},k_{\perp}+q_{\perp}) z_{\sigma}(\omega,k_{\parallel},k_{\perp}) \Bigr) \Bigr\} \Bigr] , \eqa where $\sigma = \uparrow, \downarrow$ is generalized to $\sigma = 1, 2, ..., N_{\sigma}$. Critical antiferromagnetic amplitude fluctuations give rise to the $|\omega|^{\frac{1}{2}}$ self-energy correction with a numerical constant $c_{2}$ in holon dynamics ($\Sigma_{f}^{2}$ in Fig. 3), originating from the $z = 2$ dynamics of critical fluctuations \cite{Chubukov_AF_QCP}, where $z$ is the dynamical critical exponent. On the other hand, $z = 3$ gauge fluctuations result in the $|\omega|^{\frac{2}{3}}$ self-energy correction with a numerical constant $c_{a}$ in holon dynamics ($\Sigma_{f}^{a}$ in Fig. 3) \cite{U1GT_Scaling}. Gapped ferromagnetic amplitude fluctuations do not cause any singular corrections ($\Sigma_{f}^{1}$ in Fig. 3). The polarization bubble of $\Pi^{(1)}(\boldsymbol{q},i\Omega)$ within the same patch gives rise to Landau damping for both ferromagnetic amplitude and gauge fluctuations, where $\gamma$ is a damping coefficient ($\Pi_{1}^{f}$ and $\Pi_{a}^{f}$ in Fig. 3). On the other hand, the polarization bubble of $\Pi^{(2)}(\boldsymbol{q},i\Omega)$ given by the opposite patch results in the self-energy correction of $|\omega|$ to antiferromagnetic amplitude fluctuations ($\Pi_{2}^{f}$ in Fig. 3) \cite{Chubukov_AF_QCP}, where the transverse momentum in the denominator of the Landau damping term is cut by $2 \bm{k}_{F}$, absorbed into the damping coefficient $\gamma_{2}$. One may concern that the polarization bubble at $\bm{q} = 2 \bm{k}_{F}$ will result in a more singular dependence when there exist $z = 3$ gauge fluctuations, intensively discussed in Ref. \cite{U1GT_Scaling}. However, we believe that this differs from our case, where critical antiferromagnetic amplitude fluctuations give rise to more singular self-energy corrections in holon dynamics. This issue will be more addressed below. Ferromagnetic amplitude fluctuations give rise to consistency in dynamics of transverse spin fluctuations ($\Sigma_{z}^{1}$ in Fig. 3), as discussed in the last section. They also cause the holon-spinon coupling term, which turns out to play an important role in our antiferromagnetic phase transition. Role of the spinon-gauge coupling term will be taken into account below.

Performing the Fourier transformation toward the real space, we obtain \bqa && Z = \int D f_{s\sigma} D z_{\sigma} D \phi_{n} D a \exp\Bigl[ - \int_{0}^{\beta} d \tau \int_{-\infty}^{\infty} d x \int_{-\infty}^{\infty} d y \Bigl\{ f_{s\sigma}^{\dagger} \Bigl( - i \frac{c_{2}}{N_{\sigma}} (- \partial_{\tau}^{2})^{\frac{1}{4}} - i \frac{c_{a}}{N_{\sigma}} (- \partial_{\tau}^{2})^{\frac{1}{3}} - i s v_{F} \partial_{x} - \frac{v_{F}}{2\gamma} \partial_{y}^{2} \Bigr) f_{s\sigma} \nn && + \phi_{1} \Bigl( \gamma \frac{\sqrt{- \partial_{\tau}^{2}}}{\sqrt{- \partial_{y}^{2}}} - v_{1}^{2} \partial_{y}^{2} + m_{1}^{2} \Bigr) \phi_{1} + \phi_{2} \Bigl( \gamma_{2} \sqrt{- \partial_{\tau}^{2}} - v_{2}^{2} \partial_{y}^{2} \Bigr) \phi_{2} + \frac{u_{2}}{2} \phi_{2}^{4} + a \Bigl( \gamma \frac{\sqrt{- \partial_{\tau}^{2}}}{\sqrt{- \partial_{y}^{2}}} - v_{a}^{2} \partial_{y}^{2} \Bigr) a \nn && - \frac{g_{1}}{\sqrt{N_{\sigma}}} \phi_{1} \sigma f_{s\sigma}^{\dagger} f_{s\sigma} - \frac{g_{2}}{\sqrt{N_{\sigma}}} \phi_{2} \sigma f_{s\sigma}^{\dagger} f_{-s\sigma} - \frac{e_{f}}{\sqrt{N_{\sigma}}} s v_{F} a \sigma f_{s\sigma}^{\dagger} f_{s\sigma} - \frac{g_{1} g_{z}}{N_{\sigma}} \sigma f_{s\sigma}^{\dagger} f_{s\sigma} \frac{1}{\gamma \frac{\sqrt{- \partial_{\tau}^{2}}}{\sqrt{- \partial_{y}^{2}}} - v_{1}^{2} \partial_{y}^{2} + m_{1}^{2}} (z_{\sigma'}^{\dagger} \partial_{\tau} z_{\sigma'}) \nn && - \frac{g_{z}^{2}}{2N_{\sigma}} ( z_{\sigma}^{\dagger} \partial_{\tau} z_{\sigma} ) \frac{1}{\gamma \frac{\sqrt{- \partial_{\tau}^{2}}}{\sqrt{- \partial_{y}^{2}}} - v_{1}^{2} \partial_{y}^{2} + m_{1}^{2}} ( z_{\sigma'}^{\dagger} \partial_{\tau} z_{\sigma'} ) + z_{\sigma}^{\dagger} ( - v_{z}^{2} \partial_{y}^{2} + m_{z}^{2} ) z_{\sigma} + \frac{u_{z}}{2} |z_{\sigma}|^{4} \nn && - \frac{g_{z}}{\sqrt{N_{\sigma}}} \phi_{1} z_{\sigma}^{\dagger} \partial_{\tau} z_{\sigma} - i \frac{e_{z}}{\sqrt{N_{\sigma}}} a [ z_{\sigma}^{\dagger} (\partial_{x} z_{\sigma}) - (\partial_{x} z_{\sigma}^{\dagger}) z_{\sigma}] \Bigr\} \Bigr] , \eqa where nonanalytic expressions in derivatives encode self-energy corrections. Resorting to robustness of the Fermi surface, we keep dynamics along the transverse momentum for boson excitations \cite{SSL_U1GT}. In other words, boson dynamics along $- \partial_{x}^{2}$ are not relevant.

Since ferromagnetic amplitude fluctuations are gapped, they can be neglected safely at low energies. In addition, the self-energy correction from critical antiferromagnetic amplitude fluctuations is more singular than that from gauge fluctuations at low energies, allowing us to keep $(- \partial_{\tau}^{2})^{\frac{1}{4}}$ only in the holon dynamics. As a result, we find a consistent U(1) slave spin-rotor effective field theory in terms of renormalized electrons, critical longitudinal spin fluctuations, transverse spin fluctuations, gauge fluctuations, and their interactions, where quantum corrections are taken into account in the Eliashberg approximation near antiferromagnetic quantum criticality \bqa && Z = \int D f_{s\sigma} D z_{\sigma} D \phi_{2} D a \exp\Bigl[ - \int_{0}^{\beta} d \tau \int_{-\infty}^{\infty} d x \int_{-\infty}^{\infty} d y \Bigl\{ f_{s\sigma}^{\dagger} \Bigl( - i \frac{c_{2}}{N_{\sigma}} (- \partial_{\tau}^{2})^{\frac{1}{4}} - i s v_{F} \partial_{x} - \frac{v_{F}}{2\gamma} \partial_{y}^{2} \Bigr) f_{s\sigma} \nn && + \phi_{2} \Bigl( \gamma_{2} \sqrt{- \partial_{\tau}^{2}} - v_{2}^{2} \partial_{y}^{2} \Bigr) \phi_{2} + \frac{u_{2}}{2} \phi_{2}^{4} + a \Bigl( \gamma \frac{\sqrt{- \partial_{\tau}^{2}}}{\sqrt{- \partial_{y}^{2}}} - v_{a}^{2} \partial_{y}^{2} \Bigr) a - \frac{g_{2}}{\sqrt{N_{\sigma}}} \phi_{2} \sigma f_{s\sigma}^{\dagger} f_{-s\sigma} - \frac{e_{f}}{\sqrt{N_{\sigma}}} s v_{F} a \sigma f_{s\sigma}^{\dagger} f_{s\sigma}\nn && - \frac{g_{1} g_{z}}{N_{\sigma} m_{1}^{2}} \sigma f_{s\sigma}^{\dagger} f_{s\sigma} (z_{\sigma'}^{\dagger} \partial_{\tau} z_{\sigma'}) + z_{\sigma}^{\dagger} \Bigl( - \frac{g_{z}^{2}}{2N_{\sigma} m_{1}^{2}} \partial_{\tau}^{2} - v_{z}^{2} \partial_{y}^{2} + m_{z}^{2} \Bigr) z_{\sigma} + \frac{u_{z}}{2} |z_{\sigma}|^{4} - i \frac{e_{z}}{\sqrt{N_{\sigma}}} a [ z_{\sigma}^{\dagger} (\partial_{x} z_{\sigma}) - (\partial_{x} z_{\sigma}^{\dagger}) z_{\sigma}] \Bigr\} \Bigr] . \eqa For the temporal part of the spinon dynamics, the unimodular constraint has been utilized, keeping the mass term only in the denominator at low energies. An essential aspect of this field theory is that the $|\omega|^{\frac{1}{2}}$ self-energy in the holon dynamics will affect scaling properties for all fields and interaction vertices.

\subsection{Scaling analysis in the case of $\bm{Q} \not= 2 \bm{k}_{F}$}

Equation (24) is our starting point for the renormalization group analysis, which reveals the structure of possible fixed points. Before going further, however, it is necessary to understand which critical field theory appears for antiferromagnetic quantum criticality of $\bm{Q} \not= 2 \bm{k}_{F}$, expected to clarify the difference between $\bm{Q} = 2 \bm{k}_{F}$ and $\bm{Q} \not= 2 \bm{k}_{F}$ ordering transitions. In order to understand the importance of $\bm{Q} = 2 \bm{k}_{F}$ for the emergence of local quantum criticality, we apply the U(1) slave spin-rotor formulation to the Hertz-Moriya-Millis theory and reach the following expression for an effective field theory \bqa && Z = \int D f_{s\sigma} D z_{\sigma} D \phi_{2} D \bm{a}^{T} \exp\Bigl[ - \int_{0}^{\beta} d \tau \int d^{2} \bm{r} \Bigl\{ f_{s\sigma}^{\dagger} \Bigl( - i \frac{c_{2}}{N_{\sigma}} (- \partial_{\tau}^{2})^{\frac{1}{4}} - i \bm{v}_{F}^{(s)} \cdot \bm{\nabla} \Bigr) f_{s\sigma} + \phi_{2} \Bigl( \gamma_{2} \sqrt{- \partial_{\tau}^{2}} - v_{2}^{2} \bm{\nabla}^{2} \Bigr) \phi_{2} \nn && + \frac{u_{2}}{2} \phi_{2}^{4} + \bm{a}^{T} \Bigl( \gamma \frac{\sqrt{- \partial_{\tau}^{2}}}{\sqrt{- \bm{\nabla}^{2}}} - v_{a}^{2} \bm{\nabla}^{2} \Bigr) \bm{a}^{T} + z_{\sigma}^{\dagger} \Bigl( - \frac{g_{z}^{2}}{2N_{\sigma} m_{1}^{2}} \partial_{\tau}^{2} - v_{z}^{2} \bm{\nabla}^{2} + m_{z}^{2} \Bigr) z_{\sigma} + \frac{u_{z}}{2} |z_{\sigma}|^{4} - \frac{g_{2}}{\sqrt{N_{\sigma}}} \phi_{2} \sigma f_{s\sigma}^{\dagger} f_{-s\sigma} \nn && - \frac{e_{f}}{\sqrt{N_{\sigma}}} s \bm{v}_{F} \cdot \bm{a}^{T} \sigma f_{s\sigma}^{\dagger} f_{s\sigma} - \frac{g_{1} g_{z}}{N_{\sigma} m_{1}^{2}} \sigma f_{s\sigma}^{\dagger} f_{s\sigma} (z_{\sigma'}^{\dagger} \partial_{\tau} z_{\sigma'}) - i \frac{e_{z}}{\sqrt{N_{\sigma}}} \bm{a}^{T} \cdot [ z_{\sigma}^{\dagger} ( \bm{\nabla} z_{\sigma}) - (\bm{\nabla} z_{\sigma}^{\dagger}) z_{\sigma}] \Bigr\} \Bigr] , \eqa where the Fermi velocity $\bm{v}_{F}^{(1)}$ in the patch $1$ is not parallel to the Fermi velocity $\bm{v}_{F}^{(2)}$ in the patch $2$ any more. Recall the band structure of high T$_{c}$ cuprates, shown in Ref. \cite{Metlitski_SDW}. The Fermi-surface curvature is omitted in this expression. Then, it is clear that the main difference between Eq. (24) and Eq. (25) lies in the dispersion of renormalized electrons, which turns out to play an essential role.

Performing the Fourier transformation, we obtain \bqa && Z = \int D f_{s\sigma} D z_{\sigma} D \phi_{2} D \bm{a}^{T} \exp\Bigl[ - \int_{-\infty}^{\infty} \frac{d \omega}{2\pi} \int \frac{d^{2} \bm{k}}{(2\pi)^{2}} \Bigl\{ f_{s\sigma}^{\dagger} \Bigl( - i \frac{c_{2} ~ \mbox{sgn}(\omega)}{N_{\sigma}} |\omega|^{\frac{1}{2}} - \bm{v}_{F}^{(s)} \cdot \bm{k} \Bigr) f_{s\sigma} \nn && + \phi_{2} \Bigl( \gamma_{2} |\omega| + v_{2}^{2} |\bm{k}|^{2} \Bigr) \phi_{2} + \bm{a}^{T} \Bigl( \gamma \frac{|\omega|}{|\bm{k}|} + v_{a}^{2} |\bm{k}|^{2} \Bigr) \bm{a}^{T} + z_{\sigma}^{\dagger} \Bigl( \frac{g_{z}^{2}}{2N_{\sigma} m_{1}^{2}} \omega^{2} + v_{z}^{2} |\bm{k}|^{2} + m_{z}^{2} \Bigr) z_{\sigma} \nn && - \frac{g_{1} g_{z}}{N_{\sigma} m_{1}^{2}} \int_{-\infty}^{\infty} \frac{d \omega'}{2\pi} \int \frac{d^{2} \bm{k}'}{(2\pi)^{2}} \int_{-\infty}^{\infty} \frac{d \Omega}{2\pi} \int \frac{d^{2} \bm{q}}{(2\pi)^{2}} \sigma f_{s\sigma}^{\dagger}(\omega+\Omega,\bm{k}+\bm{q}) f_{s\sigma}(\omega,\bm{k}) \Bigl(i \omega' - i \frac{\Omega}{2}\Bigr) z_{\sigma'}^{\dagger}(\omega'-\Omega,\bm{k}'-\bm{q}) z_{\sigma'}(\omega',\bm{k}') \nn && - \int_{-\infty}^{\infty} \frac{d \Omega}{2\pi} \int \frac{d^{2} \bm{q}}{(2\pi)^{2}} \Bigl( \frac{g_{2}}{\sqrt{N_{\sigma}}} \phi_{2}(\Omega,\bm{q}) \sigma f_{s\sigma}^{\dagger}(\omega+\Omega,\bm{k}+\bm{q}) f_{-s\sigma}(\omega,\bm{k}) \nn && + \frac{e_{f}}{\sqrt{N_{\sigma}}} \bm{a}^{T}(\Omega,\bm{q}) \cdot \bm{v}_{F}^{(s)} \sigma f_{s\sigma}^{\dagger}(\omega+\Omega,\bm{k}+\bm{q}) f_{s\sigma}(\omega,\bm{k}) + \frac{e_{z}}{\sqrt{N_{\sigma}}} \bm{a}^{T}(\Omega,\bm{q}) \cdot \Bigl( \bm{k} + \frac{\bm{q}}{2} \Bigr) z_{\sigma}^{\dagger}(\omega+\Omega,\bm{k}+\bm{q}) z_{\sigma}(\omega,\bm{k}) \Bigr) \Bigr\} \Bigr] . \eqa Taking the scale transformation as follows \bqa && \omega = b^{-1} \omega' , ~~~~~ \bm{k} = b^{-\frac{1}{2}} \bm{k}' , \eqa we obtain \bqa && f_{s\sigma}(\omega,\bm{k}) = b^{\frac{5}{4}} f_{s\sigma}'(\omega',\bm{k}') , ~~~~~ \phi_{2}(\omega,\bm{k}) = b^{\frac{3}{2}} \phi_{2}'(\omega',\bm{k}') , \eqa which lead their kinetic energies invariant under this scale transformation. Then, it is straightforward to verify the spin-fermion coupling constant $g_{2}$ marginal. As a result, we observe that the Hertz-Moriya-Millis theory of $\mathcal{S}_{HMM} = \int_{0}^{\beta} d \tau \int d^{2} \bm{r} \Bigl\{ f_{s\sigma}^{\dagger} \Bigl(- i \frac{c_{2}}{N_{\sigma}} (- \partial_{\tau}^{2})^{\frac{1}{4}} - i \bm{v}_{F}^{(s)} \cdot \bm{\nabla} \Bigr) f_{s\sigma} + \phi_{2} \Bigl( \gamma_{2} \sqrt{- \partial_{\tau}^{2}} - v_{2}^{2} \bm{\nabla}^{2} \Bigr) \phi_{2} - \frac{g_{2}}{\sqrt{N_{\sigma}}} \phi_{2} \sigma f_{s\sigma}^{\dagger} f_{-s\sigma} \Bigr\}$ remains to be a critical sector at this fixed point.

However, there exist more emergent excitations in the U(1) slave spin-rotor formulation, spinons and gauge fields. Unfortunately, we cannot make the kinetic energy of the gauge field and that of the spinon field invariant under this scale transformation. Assuming the Landau damping term invariant for dynamics of gauge fluctuations and the frequency-dependent term invariant for dynamics of spinons, we obtain \bqa && a(\omega,\bm{k}) = b^{\frac{5}{4}} a'(\omega',\bm{k}')  , ~~~~~ z_{\sigma}(\omega,\bm{k}) = b^{2} z_{\sigma}'(\omega',\bm{k}') . \eqa Since the Landau damping term is singular, not renormalized, and the consistency results from the renormalization in the frequency sector, we believe that these assumptions seem plausible for the renormalization group analysis. Assuming that the momentum sector in the spinon dynamics is invariant under the scale transformation, one may consider the scaling of the spinon field, given by $z_{\sigma}(\omega,\bm{k}) = b^{3/2} z_{\sigma}'(\omega',\bm{k}')$ with $v_{z} = v_{z}'$. Then, the prefactor of the frequency sector follows the scale transformation of $(g_{z}^{2}/m_{1}^{2}) = b^{1} (g_{z}^{2}/m_{1}^{2})'$, i.e., relevant, which implies condensation of spinons. The condensation of spinons allows us to neglect gauge fluctuations, where they become gapped due to Higgs mechanism. As a result, only renormalized electrons and longitudinal antiferromagnetic fluctuations appear to describe an antiferromagnetic quantum critical point, identified with the Hertz-Moriya-Millis fixed point. On the other hand, if we assume the invariance of the frequency sector in the spinon dynamics, we find that both the velocity and mass of spinons become irrelevant. Below, we focus on this scaling. The velocity of U(1) gauge fields turns out to be relevant, given by $v_{a}^{2} = b^{\frac{1}{2}} {v_{a}^{2}}'$.

Based on these scale transformations, we can determine the relevance of interaction vertices. Although the spin-fermion coupling constant $g_{2}$ is marginal, preserving the structure of the Hertz-Moriya-Millis theory, both the gauge charge of holons and the spinon-holon coupling constant turn out to be relevant, given by \bqa && e_{f} = b^{\frac{1}{4}} e_{f}', ~~~~~ \Bigl(\frac{g_{1} g_{z}}{N_{\sigma} m_{1}^{2}}\Bigr) = b^{\frac{1}{2}} \Bigl(\frac{g_{1} g_{z}}{N_{\sigma} m_{1}^{2}}\Bigr)' , \eqa but the gauge charge of spinons is irrelevant.

%
%
%
%
%
%

Incorporating these scale transformations into the effective field theory, we obtain \bqa && Z = \int D f_{s\sigma} D z_{\sigma} D \phi_{2} D \bm{a}^{T} \exp\Bigl[ - \int_{-\infty}^{\infty} \frac{d \omega}{2\pi} \int \frac{d^{2} \bm{k}}{(2\pi)^{2}} \Bigl\{ f_{s\sigma}^{\dagger} \Bigl( - i \frac{c_{2} ~ \mbox{sgn}(\omega)}{N_{\sigma}} |\omega|^{\frac{1}{2}} - \bm{v}_{F}^{(s)} \cdot \bm{k} \Bigr) f_{s\sigma} + \phi_{2} \Bigl( \gamma_{2} |\omega| + v_{2}^{2} |\bm{k}|^{2} \Bigr) \phi_{2} \nn && - \int_{-\infty}^{\infty} \frac{d \Omega}{2\pi} \int \frac{d^{2} \bm{q}}{(2\pi)^{2}} \frac{g_{2}}{\sqrt{N_{\sigma}}} \phi_{2}(\Omega,\bm{q}) \sigma f_{s\sigma}^{\dagger}(\omega+\Omega,\bm{k}+\bm{q}) f_{-s\sigma}(\omega,\bm{k}) \Bigr\} - \int_{-\infty}^{\infty} \frac{d \omega}{2\pi} \int \frac{d^{2} \bm{k}}{(2\pi)^{2}} \Bigl\{ \bm{a}^{T} \Bigl( \gamma \frac{|\omega|}{|\bm{k}|} + v_{a}^{2} |\bm{k}|^{2} \Bigr) \bm{a}^{T} \nn && + z_{\sigma}^{\dagger} \frac{g_{z}^{2}}{2N_{\sigma} m_{1}^{2}} \omega^{2} z_{\sigma} - \int_{-\infty}^{\infty} \frac{d \Omega}{2\pi} \int \frac{d^{2} \bm{q}}{(2\pi)^{2}} \frac{e_{f}}{\sqrt{N_{\sigma}}} \bm{a}^{T}(\Omega,\bm{q}) \cdot \bm{v}_{F}^{(s)} \sigma f_{s\sigma}^{\dagger}(\omega+\Omega,\bm{k}+\bm{q}) f_{s\sigma}(\omega,\bm{k}) \nn && - \frac{g_{1} g_{z}}{N_{\sigma} m_{1}^{2}} \int_{-\infty}^{\infty} \frac{d \omega'}{2\pi} \int \frac{d^{2} \bm{k}'}{(2\pi)^{2}} \int_{-\infty}^{\infty} \frac{d \Omega}{2\pi} \int \frac{d^{2} \bm{q}}{(2\pi)^{2}} \sigma f_{s\sigma}^{\dagger}(\omega+\Omega,\bm{k}+\bm{q}) f_{s\sigma}(\omega,\bm{k}) \Bigl(i \omega' - i \frac{\Omega}{2}\Bigr) z_{\sigma'}^{\dagger}(\omega'-\Omega,\bm{k}'-\bm{q}) z_{\sigma'}(\omega',\bm{k}') \Bigr\} \Bigr] . \nn \eqa Finally, we reach the following expression for an effective field theory, expected to describe an antiferromagnetic quantum criticality with $\bm{Q} \not= 2 \bm{k}_{F}$, \bqa && Z = \int D f_{s\sigma} D z_{\sigma} D \phi_{2} D \bm{a}^{T} \exp\Bigl[ - \int_{0}^{\beta} d \tau \int d^{2} \bm{r} \Bigl\{ f_{s\sigma}^{\dagger} \Bigl(- i \frac{c_{2}}{N_{\sigma}} (- \partial_{\tau}^{2})^{\frac{1}{4}} - i \bm{v}_{F}^{(s)} \cdot \bm{\nabla} \Bigr) f_{s\sigma} + \phi_{2} \Bigl( \gamma_{2} \sqrt{- \partial_{\tau}^{2}} - v_{2}^{2} \bm{\nabla}^{2} \Bigr) \phi_{2} \nn && - \frac{g_{2}}{\sqrt{N_{\sigma}}} \phi_{2} \sigma f_{s\sigma}^{\dagger} f_{-s\sigma} \Bigr\} - \int_{0}^{\beta} d \tau \int d^{2} \bm{r} \Bigl\{ \bm{a}^{T} \Bigl( \gamma \frac{\sqrt{-\partial_{\tau}^{2}}}{\sqrt{-\bm{\nabla}^{2}}} - v_{a}^{2} \bm{\nabla}^{2} \Bigr) \bm{a}^{T} - \frac{e_{f}}{\sqrt{N_{\sigma}}} \bm{a}^{T} \cdot \bm{v}_{F}^{(s)} \sigma f_{s\sigma}^{\dagger} f_{s\sigma} + |\partial_{\tau} z_{\sigma}(\tau,x,y)|^{2} \nn && - \frac{g_{1} g_{z}}{N_{\sigma} m_{1}^{2}} \sigma f_{s\sigma}^{\dagger} f_{s\sigma} (z_{\sigma'}^{\dagger} \partial_{\tau} z_{\sigma'})  \Bigr\} \Bigr] . \eqa This expression is quite interesting due to the following reasoning. First, we note that the Hertz-Moriya-Millis theory of $\mathcal{S}_{HMM} = \int_{0}^{\beta} d \tau \int d^{2} \bm{r} \Bigl\{ f_{s\sigma}^{\dagger} \Bigl(- i \frac{c_{2}}{N_{\sigma}} (- \partial_{\tau}^{2})^{\frac{1}{4}} - i \bm{v}_{F}^{(s)} \cdot \bm{\nabla} \Bigr) f_{s\sigma} + \phi_{2} \Bigl( \gamma_{2} \sqrt{- \partial_{\tau}^{2}} - v_{2}^{2} \bm{\nabla}^{2} \Bigr) \phi_{2} - \frac{g_{2}}{\sqrt{N_{\sigma}}} \phi_{2} \sigma f_{s\sigma}^{\dagger} f_{-s\sigma} \Bigr\}$, regarded to be the standard critical field theory for antiferromagnetic quantum criticality, becomes modified in the strong coupling approach of the U(1) slave spin-rotor representation. In particular, the effective interaction $e_{f}$ between holons and U(1) gauge fluctuations is relevant, originating from the fact that the self-energy correction for holons results from the $z = 2$ critical dynamics of longitudinal spin fluctuations, not matched with the $z = 3$ dynamics of U(1) gauge fluctuations. In addition, such renormalized electrons interact with emergent localized transverse spin fluctuations strongly. We also point out that the propagating velocity of U(1) gauge fields is relevant, modifying the dispersion at low energies. We suspect the possibility that the gauge dynamics becomes renormalized into $\sqrt{- \bm{\nabla}^{2}}$, turning from $z = 3$ to $z = 2$. Then, not only $g_{2}$ but also $e_{f}$ becomes marginal. Even in this scenario, the spinon-holon interaction is still relevant. As a result, we conclude that the fixed point described by the U(1) slave spin-rotor theory should be distinguished from the Hertz-Moriya-Millis fixed point. We suggest to call it U(1) slave spin-rotor fixed point for antiferromagnetic quantum criticality with $\bm{Q} \not= 2 \bm{k}_{F}$.

\begin{figure}[t]
\includegraphics[width=0.8\textwidth]{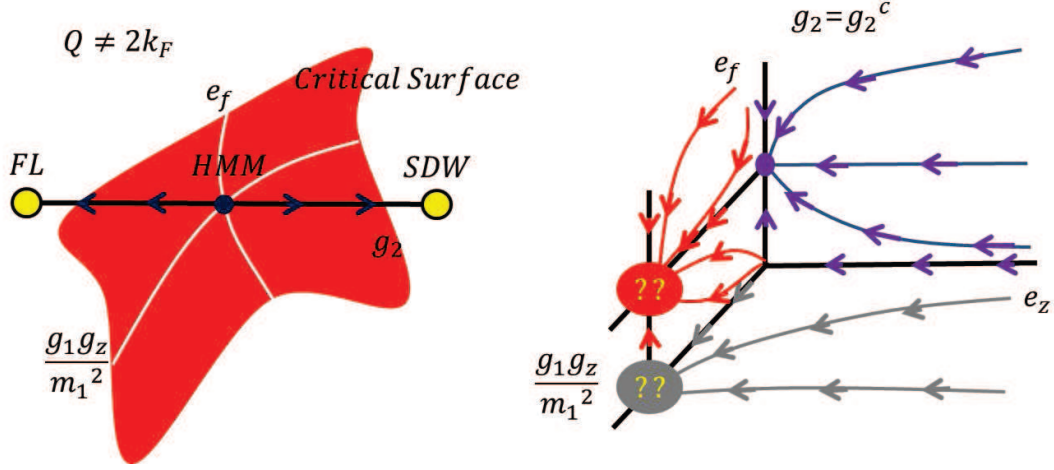}
\caption{A schematic diagram for renormalization group flows in an antiferromagnetic quantum phase transition with $\bm{Q} \not= 2 \bm{k}_{F}$. The U(1) slave spin-rotor theory suggests possible existence of another fixed point beyond the Hertz-Moriya-Millis theory on the ``Hertz-Moriya-Millis" critical surface. When spinons are forced to condense, the Hertz-Moriya-Millis fixed point would be realized. On the other hand, if dynamics of such transverse spin fluctuations becomes localized, we expect to reach the U(1) slave spin-rotor fixed point, differentiated from the Hertz-Moriya-Millis. In this study the nature of the U(1) slave spin-rotor fixed point is not clarified.} \label{RG_Flow_Qnot2kF}
\end{figure}

Figure 4 shows a schematic diagram for renormalization group flows in an antiferromagnetic quantum phase transition with $\bm{Q} \not= 2 \bm{k}_{F}$. An essential feature is that the U(1) slave spin-rotor theory allows another fixed point, not described by the Hertz-Moriya-Millis theory as discussed above. One may define a critical surface given by $g_{2} = g_{2}^{c}$ as usual. The U(1) slave spin-rotor theory gives rise to a fine structure on this critical surface, where two types of effective interactions denoted by $e_{f}$ (between holons and gauge fluctuations) and $\frac{g_{1} g_{z}}{N_{\sigma} m_{1}^{2}}$ (between renormalized electrons and transverse spin fluctuations) separate the Hertz-Moriya-Millis fixed point from the other. When spinons are forced to condense, the Hertz-Moriya-Millis fixed point would be realized, as discussed before. On the other hand, if dynamics of such transverse spin fluctuations becomes localized, we reach the U(1) slave spin-rotor fixed point. Unfortunately, we are not in a position, resolving this problem, which means when the U(1) slave spin-rotor fixed point occurs. We also point out that the nature of the U(1) slave spin-rotor fixed point is not clarified in this study. Although we speculate that there appears screening to result in a finite-coupling fixed point given by $\Bigl(\frac{g_{1} g_{z}}{N_{\sigma} m_{1}^{2}}\Bigr)_{c}$, we do not exclude a possible run-away flow along this direction.

\subsection{Scaling analysis I}

We return to our original problem on antiferromagnetic quantum criticality with $\bm{Q} = 2 \bm{k}_{F}$. Performing the Fourier transformation for our effective field theory Eq. (24), we obtain \bqa && Z = \int D f_{s\sigma} D z_{\sigma} D \phi_{2} D a \exp\Bigl[ - \int_{-\infty}^{\infty} \frac{d \omega}{2\pi} \int_{-\infty}^{\infty} \frac{d k_{\parallel}}{2\pi} \int_{-\infty}^{\infty} \frac{d k_{\perp}}{2\pi} \Bigl\{ f_{s\sigma}^{\dagger} \Bigl( - i \frac{c_{2} ~ \mbox{sgn}(\omega)}{N_{\sigma}} |\omega|^{\frac{1}{2}} - s v_{F} k_{\parallel} - \frac{v_{F}}{2\gamma} k_{\perp}^{2} \Bigr) f_{s\sigma} \nn && + \phi_{2} \Bigl( \gamma_{2} |\omega| + v_{2}^{2} k_{\perp}^{2} \Bigr) \phi_{2} + a \Bigl( \gamma \frac{|\omega|}{|k_{\perp}|} + v_{a}^{2} k_{\perp}^{2} \Bigr) a + z_{\sigma}^{\dagger} \Bigl( \frac{g_{z}^{2}}{2N_{\sigma} m_{1}^{2}} \omega^{2} + v_{z}^{2} k_{\perp}^{2} + m_{z}^{2} \Bigr) z_{\sigma} \nn && - \frac{g_{1} g_{z}}{N_{\sigma} m_{1}^{2}} \int_{-\infty}^{\infty} \frac{d \omega'}{2\pi} \int_{-\infty}^{\infty} \frac{d k_{\parallel}'}{2\pi} \int_{-\infty}^{\infty}\frac{d k_{\perp}}{2\pi} \int_{-\infty}^{\infty} \frac{d \Omega}{2\pi} \int_{-\infty}^{\infty} \frac{d q_{\parallel}}{2\pi} \int_{-\infty}^{\infty} \frac{d q_{\perp}}{2\pi} \sigma f_{s\sigma}^{\dagger}(\omega+\Omega,k_{\parallel}+q_{\parallel},k_{\perp}+q_{\perp}) f_{s\sigma}(\omega,k_{\parallel},k_{\perp}) \nn && \Bigl(i \omega' - i \frac{\Omega}{2}\Bigr) z_{\sigma'}^{\dagger}(\omega'-\Omega,k_{\parallel}'-q_{\parallel},k_{\perp}'-q_{\perp}) z_{\sigma'}(\omega',k_{\parallel}',k_{\perp}') \nn && - \int_{-\infty}^{\infty} \frac{d \Omega}{2\pi} \int_{-\infty}^{\infty} \frac{d q_{\parallel}}{2\pi} \int_{-\infty}^{\infty} \frac{d q_{\perp}}{2\pi} \Bigl( \frac{g_{2}}{\sqrt{N_{\sigma}}} \phi_{2}(\Omega,q_{\parallel},q_{\perp}) \sigma f_{s\sigma}^{\dagger}(\omega+\Omega,k_{\parallel}+q_{\parallel},k_{\perp}+q_{\perp}) f_{-s\sigma}(\omega,k_{\parallel},k_{\perp}) \nn && + \frac{e_{f}}{\sqrt{N_{\sigma}}} s a(\Omega,q_{\parallel},q_{\perp}) v_{F} \sigma f_{s\sigma}^{\dagger}(\omega+\Omega,k_{\parallel}+q_{\parallel},k_{\perp}+q_{\perp}) f_{s\sigma}(\omega,k_{\parallel},k_{\perp}) \nn && + \frac{e_{z}}{\sqrt{N_{\sigma}}} a(\Omega,q_{\parallel},q_{\perp}) \Bigl( k_{\parallel} + \frac{q_{\parallel}}{2} \Bigr) z_{\sigma}^{\dagger}(\omega+\Omega,k_{\parallel}+q_{\parallel},k_{\perp}+q_{\perp}) z_{\sigma}(\omega,k_{\parallel},k_{\perp}) \Bigr) \Bigr\} \Bigr] . \eqa

Assuming the robustness of fermion dynamics, we introduce the scale transformation of \bqa && \omega = b^{-1} \omega' , ~~~~~ k_{\parallel} = b^{-\frac{1}{2}} k_{\parallel}' , ~~~~~ k_{\perp} = b^{-\frac{1}{4}} k_{\perp}' , \eqa which leads the holon's renormalized kinetic energy invariant under the transformation of \bqa && f_{s\sigma}(\omega,k_{\parallel},k_{\perp}) = b^{\frac{9}{8}} f_{s\sigma}'(\omega',k_{\parallel}',k_{\perp}') . \eqa On the other hand, the kinetic energy of critical longitudinal spin fluctuations cannot be invariant under this scale transformation. It is natural to assume the invariance of the Landau damping term under the scale transformation, resulting in \bqa && \phi_{2}(\omega,k_{\parallel},k_{\perp}) = b^{\frac{11}{8}} \phi_{2}'(\omega',k_{\parallel}',k_{\perp}') . \eqa Then, we find \bqa && v_{2} = b^{- \frac{1}{4}} v_{2}' , \eqa which turns out to be irrelevant at low energies. In the same way it is natural to keep the invariance of the Landau damping term under the scale transformation for dynamics of gauge fluctuations, which gives \bqa && a(\omega,k_{\parallel},k_{\perp}) = b^{\frac{5}{4}} a'(\omega',k_{\parallel}',k_{\perp}') . \eqa Then, the velocity of gauge fluctuations also turns out to be irrelevant. For the scale transformation in transverse spin fluctuations, we keep the invariance of the frequency term. If we consider the scale invariance of the momentum term in dynamics of transverse spin fluctuations, the frequency term turns out to be relevant. Then, spinon dynamics becomes static at low energies since only the zero-frequency sector is allowed, as discussed before. The condensation of spinons leads us to return back to the Hertz-Moriya-Millis description. However, we find that the spin-fermion coupling $g_{2}$ becomes irrelevant, which means that dynamics of localized longitudinal antiferromagnetic fluctuations is decoupled from that of renormalized electrons in the zero-temperature limit. Of course, there must be quantum corrections at finite temperatures due to effective interactions between renormalized electrons and such localized longitudinal antiferromagnetic fluctuations, given by $\mathcal{S}_{HMM} = \int_{0}^{\beta} d \tau \int d^{2} \bm{r} \Bigl\{ f_{s\sigma}^{\dagger} \Bigl(- i \frac{c_{2}}{N_{\sigma}} (- \partial_{\tau}^{2})^{\frac{1}{4}} - i \bm{v}_{F}^{(s)} \cdot \bm{\nabla} \Bigr) f_{s\sigma} + \gamma_{2} \phi_{2} \sqrt{- \partial_{\tau}^{2}} \phi_{2} - \frac{g_{2}}{\sqrt{N_{\sigma}}} \phi_{2} \sigma f_{s\sigma}^{\dagger} f_{-s\sigma} \Bigr\}$. We do not consider this local quantum criticality any more in this study. We focus on the case, where spinons do not condense. The scale invariance of the frequency part is achieved by \bqa && z_{\sigma}(\omega,k_{\parallel},k_{\perp}) = b^{\frac{15}{8}} z_{\sigma}'(\omega',k_{\parallel}',k_{\perp}') . \eqa As a result, we find that both the velocity and mass of spinons become irrelevant, which decrease to vanish at low energies.

It is straightforward to show that both $g_{2}$ and $e_{z}$ are irrelevant. On the other hand, $e_{f}$ is marginal. The interaction vertex between renormalized electrons and transverse spin fluctuations turns out to be relevant, given by \bqa && \Bigl(\frac{g_{1} g_{z}}{N_{\sigma} m_{1}^{2}}\Bigr) = b^{\frac{1}{4}} \Bigl(\frac{g_{1} g_{z}}{N_{\sigma} m_{1}^{2}}\Bigr)' . \eqa As a result, we reach the following expression for an effective field theory at this fixed point \bqa && Z = \int D f_{s\sigma} D z_{\sigma} D a \exp\Bigl[ - \int_{-\infty}^{\infty} \frac{d \omega}{2\pi} \int_{-\infty}^{\infty} \frac{d k_{\parallel}}{2\pi} \int_{-\infty}^{\infty} \frac{d k_{\perp}}{2\pi} \Bigl\{ f_{s\sigma}^{\dagger}(\omega,k_{\parallel},k_{\perp}) \Bigl( - i \frac{c_{2} \mbox{sgn}(\omega)}{N_{\sigma}} |\omega|^{\frac{1}{2}} - s v_{F} k_{\parallel} \nn && - \frac{v_{F}}{2\gamma} k_{\perp}^{2} \Bigr) f_{s\sigma}(\omega,k_{\parallel},k_{\perp}) + a(\omega,k_{\parallel},k_{\perp}) \gamma \frac{|\omega|}{|k_{\perp}|} a(-\omega,-k_{\parallel},-k_{\perp}) + z_{\sigma}^{\dagger}(\omega,k_{\parallel},k_{\perp}) \omega^{2} z_{\sigma}(\omega,k_{\parallel},k_{\perp}) \nn && - \int_{-\infty}^{\infty} \frac{d \Omega}{2\pi} \int_{-\infty}^{\infty} \frac{d q_{\parallel}}{2\pi} \int_{-\infty}^{\infty} \frac{d q_{\perp}}{2\pi} \frac{e_{f}}{\sqrt{N_{\sigma}}} s a(\Omega,q_{\parallel},q_{\perp}) v_{F} \sigma f_{s\sigma}^{\dagger}(\omega+\Omega,k_{\parallel}+q_{\parallel},k_{\perp}+q_{\perp}) f_{s\sigma}(\omega,k_{\parallel},k_{\perp}) \nn && - \frac{g_{1} g_{z}}{N_{\sigma} m_{1}^{2}} \int_{-\infty}^{\infty} \frac{d \omega'}{2\pi} \int_{-\infty}^{\infty} \frac{d k_{\parallel}'}{2\pi} \int_{-\infty}^{\infty}\frac{d k_{\perp}}{2\pi} \int_{-\infty}^{\infty} \frac{d \Omega}{2\pi} \int_{-\infty}^{\infty} \frac{d q_{\parallel}}{2\pi} \int_{-\infty}^{\infty} \frac{d q_{\perp}}{2\pi} \sigma f_{s\sigma}^{\dagger}(\omega+\Omega,k_{\parallel}+q_{\parallel},k_{\perp}+q_{\perp}) f_{s\sigma}(\omega,k_{\parallel},k_{\perp}) \nn && \Bigl(i \omega' - i \frac{\Omega}{2}\Bigr) z_{\sigma'}^{\dagger}(\omega'-\Omega,k_{\parallel}'-q_{\parallel},k_{\perp}'-q_{\perp}) z_{\sigma'}(\omega',k_{\parallel}',k_{\perp}') \Bigr\} \Bigr] . \eqa
Performing the Fourier transformation, we express the above effective field theory as follows \bqa && Z = \int D f_{s\sigma} D z_{\sigma} D a \exp\Bigl[ - \int_{0}^{\beta} d \tau \int_{-\infty}^{\infty} d x \int_{-\infty}^{\infty} d y \Bigl\{ f_{s\sigma}^{\dagger}(\tau,x,y) \Bigl(- i \frac{c_{2}}{N_{\sigma}} (- \partial_{\tau}^{2})^{\frac{1}{4}} - i s v_{F} \partial_{x} - \frac{v_{F}}{2\gamma} \partial_{y}^{2} \Bigr) f_{s\sigma}(\tau,x,y) \nn && + a(\tau,x,y) \gamma \frac{\sqrt{-\partial_{\tau}^{2}}}{\sqrt{-\partial_{y}^{2}}} a(\tau,x,y) - \frac{e_{f}}{\sqrt{N_{\sigma}}} s a(\tau,x,y) v_{F} \sigma f_{s\sigma}^{\dagger}(\tau,x,y) f_{s\sigma}(\tau,x,y) + |\partial_{\tau} z_{\sigma}(\tau,x,y)|^{2} \nn && - \frac{g_{1} g_{z}}{N_{\sigma} m_{1}^{2}} \sigma f_{s\sigma}^{\dagger} f_{s\sigma} (z_{\sigma'}^{\dagger} \partial_{\tau} z_{\sigma'})  \Bigr\} \Bigr] . \eqa It is interesting to notice that dynamics of gauge fluctuations becomes local, which means that they do not propagate in space. This locality originates from the fact that the renormalization in fermion dynamics results from $z = 2$ critical dynamics of longitudinal spin fluctuations while dynamics of gauge fluctuations is given by $z = 3$. Although it may be possible to reach a fixed point of $\Bigl(\frac{g_{1} g_{z}}{N_{\sigma} m_{1}^{2}}\Bigr)_{c}$, we fail to find the corresponding critical field theory within the Eliashberg approximation.
%
%

\subsection{Scaling analysis II}

We emphasize that Eq. (42) is not a critical field theory since the holon-spinon coupling term is relevant. In order to find a fixed-point theory within the one-loop level, we consider another scale transformation, given by \bqa && \omega = b^{-1} \omega' , ~~~~~ k_{\parallel} = b^{-\frac{1}{2}} k_{\parallel}' , ~~~~~ k_{\perp} = k_{\perp}' , \eqa where the transverse momentum along the Fermi surface does not change under the scale transformation. We speculate that it is hidden in this scale transformation one dimensional physics associated with an emergent localized magnetic moment.

Following the same procedure, we find that all field variables change as follows \bqa && f_{s\sigma}(\omega,k_{\parallel},k_{\perp}) = b^{1} f_{s\sigma}'(\omega',k_{\parallel}',k_{\perp}') , ~~~~~ \phi(\omega,k_{\parallel},k_{\perp}) = b^{\frac{5}{4}} \phi'(\omega',k_{\parallel}',k_{\perp}') , \nn && a(\omega,k_{\parallel},k_{\perp}) = b^{\frac{5}{4}} a'(\omega',k_{\parallel}',k_{\perp}') , ~~~~~ z_{\sigma}(\omega,k_{\parallel},k_{\perp}) = b^{\frac{7}{4}} z_{\sigma}'(\omega',k_{\parallel}',k_{\perp}') \eqa under the effectively one-dimensional scale transformation. It turns out that all interaction vertices become irrelevant except for the spinon-holon coupling term, which is marginal. As a result, we reach the following expression for a critical field theory
%
%
\bqa && Z = \int D f_{s\sigma} D z_{\sigma} \exp\Bigl[ - \int_{-\infty}^{\infty} \frac{d \omega}{2\pi} \int_{-\infty}^{\infty} \frac{d k_{\parallel}}{2\pi} \int_{-\infty}^{\infty} \frac{d k_{\perp}}{2\pi} \Bigl\{ f_{s\sigma}^{\dagger}(\omega,k_{\parallel},k_{\perp}) \Bigl( - i \frac{c_{\phi} \mbox{sgn}(\omega)}{N_{\sigma}} |\omega|^{\frac{1}{2}} - s v_{F} k_{\parallel} \Bigr) f_{s\sigma}(\omega,k_{\parallel},k_{\perp}) \nn && + z_{\sigma}^{\dagger}(\omega,k_{\parallel},k_{\perp}) \omega^{2} z_{\sigma}(\omega,k_{\parallel},k_{\perp}) - \frac{g_{1} g_{z}}{N_{\sigma} m_{1}^{2}} \int_{-\infty}^{\infty} \frac{d \omega'}{2\pi} \int_{-\infty}^{\infty} \frac{d k_{\parallel}'}{2\pi} \int_{-\infty}^{\infty} \frac{d k_{\perp}'}{2\pi} \int_{-\infty}^{\infty} \frac{d \Omega}{2\pi} \int_{-\infty}^{\infty} \frac{d q_{\parallel}}{2\pi} \int_{-\infty}^{\infty} \frac{d q_{\perp}}{2\pi} \nn && \sigma f_{s\sigma}^{\dagger}(\omega+\Omega,k_{\parallel}+q_{\parallel},k_{\perp}+q_{\perp}) f_{s\sigma}(\omega,k_{\parallel},k_{\perp}) \Bigl(i \omega' - i \frac{\Omega}{2}\Bigr) z_{\sigma'}^{\dagger}(\omega'-\Omega,k_{\parallel}'-q_{\parallel},k_{\perp}'-q_{\perp}) z_{\sigma'}(\omega',k_{\parallel}',k_{\perp}') \Bigr\} \Bigr] . \eqa Performing the Fourier transformation, we obtain a critical field theory given by \bqa && Z = \int D f_{s\sigma} D z_{\sigma} \exp\Bigl[ - \int_{0}^{\beta} d \tau \int_{-\infty}^{\infty} d x \Bigl\{ f_{s\sigma}^{\dagger}(\tau,x) \Bigl(- i \frac{c_{\phi}}{N_{\sigma}} (- \partial_{\tau}^{2})^{\frac{1}{4}} - i s v_{F} \partial_{x} \Bigr) f_{s\sigma}(\tau,x) \nn && + |\partial_{\tau} z_{\sigma}(\tau,x)|^{2} - \frac{g_{1} g_{z}}{N_{\sigma} m_{1}^{2}} \sigma f_{s\sigma}^{\dagger}(\tau,x) f_{s\sigma}(\tau,x) [z_{\sigma'}^{\dagger}(\tau,x) \partial_{\tau} z_{\sigma'}(\tau,x)] \Bigr\} \Bigr] , \eqa evaluated within the Eliashberg approximation.

This critical field theory describes dynamics of renormalized electrons interacting with locally critical transverse spin fluctuations at the antiferromagnetic quantum critical point with $\bm{Q} = 2 \bm{k}_{F}$. We would like to point out that not only longitudinal spin fluctuations but also transverse gauge fluctuations have been introduced to renormalize dynamics of both renormalized electrons and transverse spin fluctuations in the Eliashberg approximation and their renormalized interactions do not play any roles more in dynamics of both renormalized electrons and transverse spin fluctuations at this ``tree" level. Considering the fact that dynamics of transverse spin fluctuations becomes localized, we interpret that localized magnetic moments appear to be critical at this antiferromagnetic quantum critical point.



\begin{figure}[t]
\includegraphics[width=0.8\textwidth]{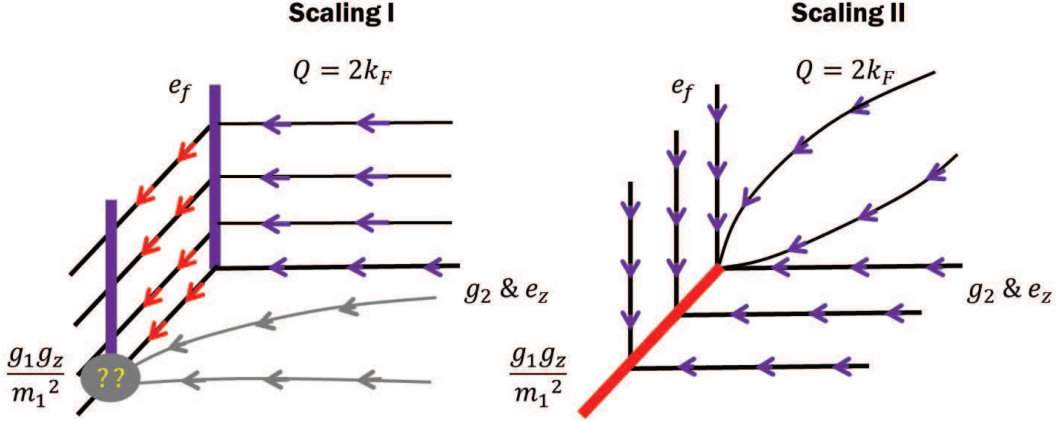}
\caption{A schematic diagram for renormalization group flows in an antiferromagnetic quantum phase transition with $\bm{Q} = 2 \bm{k}_{F}$. The U(1) slave spin-rotor formulation allows three possible fixed points. The first fixed point is realized by the condensation of spinons, described by the ``Hertz-Moriya-Millis" scenario, but dynamics of longitudinal antiferromagnetic fluctuations becomes localized at this fixed point. Since the spin-fermion coupling turns out to be irrelevant, not shown here, the local quantum criticality given by longitudinal antiferromagnetic fluctuations is decoupled from dynamics of renormalized electrons in the low energy limit. The second is described by Eq. (42), but not a critical field theory, where dynamics of all bosonic excitations of gauge fields, longitudinal antiferromagnetic fluctuations, and transverse spin fluctuations become localized and $g_{2}$ is irrelevant, but $\frac{g_{1} g_{z}}{N_{\sigma} m_{1}^{2}}$ is relevant while $e_{f}$ is marginal (Left). The last fixed point is well defined, described by the critical field theory of Eq. (46) in terms of renormalized electrons and localized transverse spin fluctuations. All interaction vertices turn out to be irrelevant except for the holon-spinon coupling $\frac{g_{1} g_{z}}{N_{\sigma} m_{1}^{2}}$, marginal in the one-loop level (Right).} \label{RG_Flow_Q2kF}
\end{figure}

Figure 5 shows a schematic diagram for renormalization group flows in an antiferromagnetic quantum phase transition with $\bm{Q} = 2 \bm{k}_{F}$. Here, we took into account three possibilities, based on the U(1) slave spin-rotor formulation. The first case is given by the condensation of spinons, reproducing the Hertz-Moriya-Millis description. However, it turns out that dynamics of longitudinal antiferromagnetic fluctuations becomes localized and the spin-fermion coupling is irrelevant in the low energy limit. As a result, we reach the ``Hertz-Moriya-Millis" fixed point of $\bm{Q} = 2 \bm{k}_{F}$, described by $\mathcal{S}_{HMM} = \int_{0}^{\beta} d \tau \int d^{2} \bm{r} \Bigl\{ f_{s\sigma}^{\dagger} \Bigl(- i \frac{c_{2}}{N_{\sigma}} (- \partial_{\tau}^{2})^{\frac{1}{4}} - i \bm{v}_{F}^{(s)} \cdot \bm{\nabla} \Bigr) f_{s\sigma} + \gamma_{2} \phi_{2} \sqrt{- \partial_{\tau}^{2}} \phi_{2} - \frac{g_{2}}{\sqrt{N_{\sigma}}} \phi_{2} \sigma f_{s\sigma}^{\dagger} f_{-s\sigma} \Bigr\}$, where the spin-fermion coupling is expected to govern anomalous scaling at finite temperatures. Of course, this ``Hertz-Moriya-Millis" fixed point differs from that of $\bm{Q} \not= 2 \bm{k}_{F}$. The second fixed point is described by Eq. (42) but not a critical field theory, where dynamics of all bosonic excitations of gauge fields, longitudinal antiferromagnetic fluctuations, and transverse spin fluctuations become localized and $g_{2}$ is irrelevant, but $\frac{g_{1} g_{z}}{N_{\sigma} m_{1}^{2}}$ is relevant while $e_{f}$ is marginal. Although we speculate the existence of a finite coupling fixed point for $\frac{g_{1} g_{z}}{N_{\sigma} m_{1}^{2}}$, it is not completely clarified yet. The last is described by Eq. (46), a critical field theory in terms of renormalized electrons and localized transverse spin fluctuations. All interaction vertices turn out to be irrelevant except for the holon-spinon coupling $\frac{g_{1} g_{z}}{N_{\sigma} m_{1}^{2}}$, marginal in the one-loop level. This well defined fixed point is identified with antiferromagnetic local quantum criticality of $\bm{Q} = 2 \bm{k}_{F}$, which may be regarded to be our main result.


It is straightforward to extend the present analysis into the three dimensional case, where the fermion self-energy is proportional to $|\omega|$ linearly. It is rather unexpected to find that not only the scaling ansatz II but also the scaling ansatz I gives rise to exactly the same critical field theory, that is, Eq. (46), although it is trivial to see that the scaling ansatz II preserves the critical field theory of Eq. (46) in three dimensions. This three dimensional demonstration suggests the robustness of the local critical field theory of Eq. (46). Below, we confirm this robustness, assuming possible singular corrections in the self-energy of longitudinal antiferromagnetic fluctuations.

\subsection{Consideration on possible singular physics of the $2\bm{k}_{F}$ susceptibility}

One may concern that the self-energy correction $\Pi^{(2)}(\bm{q} \approx 2 \bm{k}_{F},i\Omega)$ for longitudinal spin (antiferromagnetic amplitude) fluctuations contains more singular dependence in frequency than the form of Landau damping since there exist $z = 3$ transverse gauge fluctuations. Actually, the $2 \bm{k}_{F}$ antiferromagnetic-ordering transition has been investigated in the U(1) spin-liquid state, described by fermionic spinons strongly coupled to U(1) gauge fluctuations of $z = 3$ \cite{U1GT_SDW}. This study addressed that the $2\bm{k}_{F}$ susceptibility is possible to show a divergent behavior at low energies, expected to modify the dynamics of spin fluctuations in the Hertz-Moriya-Millis theory.

First of all, we point out that an antiferromagnetic transition from a Fermi-liquid state is being considered instead of the U(1) spin-liquid phase. Of course, the $2 \bm{k}_{F}$ susceptibility does not show such a singular behavior in the Fermi-liquid state. Even if so, one may suspect that the emergence of U(1) gauge fluctuations, which arises from the strong-coupling approach for the antiferromagnetic quantum critical point, can cause the similar singular behavior for the $2 \bm{k}_{F}$ susceptibility. However, we would like to argue that the correct way of renormalization for the dynamics of longitudinal spin fluctuations may differ from that in the U(1) spin-liquid state, where both U(1) gauge fluctuations and critical longitudinal spin fluctuations should be taken into account on equal footing. In this respect we performed the scaling analysis for an effective field theory with quantum corrections in the Eliashberg approximation, showing that the effective interaction between holons and U(1) gauge fluctuations turns out to be irrelevant for the fixed point given by the scale transformation of Eqs. (43) and (44). This implies that singular vertex corrections for the $2 \bm{k}_{F}$ susceptibility given by U(1) gauge fluctuations may not exist at this fixed point, suggesting the local quantum criticality.

In this subsection we extend the scaling analysis for the $z = 2$ critical dynamics of longitudinal spin fluctuations to a general $z$. This consideration is meaningful since more elaborate treatments for renormalizations may change the $z = 2$ critical dynamics into the other. Modifying the Landau-damping term $|\omega|$ into $|\omega|^{\frac{2}{z}}$, we obtain an effective field theory with quantum corrections in the one-loop level \bqa && Z = \int D f_{s\sigma} D z_{\sigma} D \phi_{2} D a \exp\Bigl[ - \int_{-\infty}^{\infty} \frac{d \omega}{2\pi} \int_{-\infty}^{\infty} \frac{d k_{\parallel}}{2\pi} \int_{-\infty}^{\infty} \frac{d k_{\perp}}{2\pi} \Bigl\{ f_{s\sigma}^{\dagger} \Bigl( - i \frac{c_{2} ~ \mbox{sgn}(\omega)}{N_{\sigma}} |\omega|^{\frac{z-1}{z}} - s v_{F} k_{\parallel} - \frac{v_{F}}{2\gamma} k_{\perp}^{2} \Bigr) f_{s\sigma} \nn && + \phi_{2} \Bigl( \gamma_{2} |\omega|^{\frac{2}{z}} + v_{2}^{2} k_{\perp}^{2} \Bigr) \phi_{2} + a \Bigl( \gamma \frac{|\omega|}{|k_{\perp}|} + v_{a}^{2} k_{\perp}^{2} \Bigr) a + z_{\sigma}^{\dagger} \Bigl( \frac{g_{z}^{2}}{2N_{\sigma} m_{1}^{2}} \omega^{2} + v_{z}^{2} k_{\perp}^{2} + m_{z}^{2} \Bigr) z_{\sigma} \nn && - \frac{g_{1} g_{z}}{N_{\sigma} m_{1}^{2}} \int_{-\infty}^{\infty} \frac{d \omega'}{2\pi} \int_{-\infty}^{\infty} \frac{d k_{\parallel}'}{2\pi} \int_{-\infty}^{\infty}\frac{d k_{\perp}}{2\pi} \int_{-\infty}^{\infty} \frac{d \Omega}{2\pi} \int_{-\infty}^{\infty} \frac{d q_{\parallel}}{2\pi} \int_{-\infty}^{\infty} \frac{d q_{\perp}}{2\pi} \sigma f_{s\sigma}^{\dagger}(\omega+\Omega,k_{\parallel}+q_{\parallel},k_{\perp}+q_{\perp}) f_{s\sigma}(\omega,k_{\parallel},k_{\perp}) \nn && \Bigl(i \omega' - i \frac{\Omega}{2}\Bigr) z_{\sigma'}^{\dagger}(\omega'-\Omega,k_{\parallel}'-q_{\parallel},k_{\perp}'-q_{\perp}) z_{\sigma'}(\omega',k_{\parallel}',k_{\perp}') \nn && - \int_{-\infty}^{\infty} \frac{d \Omega}{2\pi} \int_{-\infty}^{\infty} \frac{d q_{\parallel}}{2\pi} \int_{-\infty}^{\infty} \frac{d q_{\perp}}{2\pi} \Bigl( \frac{g_{2}}{\sqrt{N_{\sigma}}} \phi_{2}(\Omega,q_{\parallel},q_{\perp}) \sigma f_{s\sigma}^{\dagger}(\omega+\Omega,k_{\parallel}+q_{\parallel},k_{\perp}+q_{\perp}) f_{-s\sigma}(\omega,k_{\parallel},k_{\perp}) \nn && + \frac{e_{f}}{\sqrt{N_{\sigma}}} s a(\Omega,q_{\parallel},q_{\perp}) v_{F} \sigma f_{s\sigma}^{\dagger}(\omega+\Omega,k_{\parallel}+q_{\parallel},k_{\perp}+q_{\perp}) f_{s\sigma}(\omega,k_{\parallel},k_{\perp}) \nn && + \frac{e_{z}}{\sqrt{N_{\sigma}}} a(\Omega,q_{\parallel},q_{\perp}) \Bigl( k_{\parallel} + \frac{q_{\parallel}}{2} \Bigr) z_{\sigma}^{\dagger}(\omega+\Omega,k_{\parallel}+q_{\parallel},k_{\perp}+q_{\perp}) z_{\sigma}(\omega,k_{\parallel},k_{\perp}) \Bigr) \Bigr\} \Bigr] , \eqa where the holon self-energy has been also changed from $|\omega|^{1/2}$ to $|\omega|^{\frac{z-1}{z}}$. If one inserts $z = 3$, he has $|\omega|^{2/3}$ as expected \cite{U1GT_Scaling}.

First, we consider the scale transformation \bqa && \omega = b^{-1} \omega' , ~~~~~ k_{\parallel} = b^{-\frac{z-1}{z}} k_{\parallel}' , ~~~~~ k_{\perp} = b^{-\frac{z-1}{2z}} k_{\perp}' , \eqa where $z \geq 1$ is required. Then, consistent transformations for all field variables are given by \bqa && f_{s\sigma}(\omega,k_{\parallel},k_{\perp}) = b^{\frac{7z-5}{4z}} f_{s\sigma}'(\omega',k_{\parallel}',k_{\perp}') , ~~~~~ \phi_{2}(\omega,k_{\parallel},k_{\perp}) = b^{\frac{5z+1}{4z}} \phi_{2}'(\omega',k_{\parallel}',k_{\perp}') , \nn && a(\omega,k_{\parallel},k_{\perp}) = b^{\frac{3z-1}{2z}} a'(\omega',k_{\parallel}',k_{\perp}') , ~~~~~ z_{\sigma}(\omega,k_{\parallel},k_{\perp}) = b^{\frac{9z-3}{4z}} z_{\sigma}'(\omega',k_{\parallel}',k_{\perp}') , \eqa following the same strategy of the scaling analysis I. As a result, we obtain \bqa && v_{2}^{2} = b^{- \frac{3-z}{z}} {v_{2}^{2}}' , ~~~~~ v_{a}^{2} = b^{-\frac{1}{z}} {v_{a}^{2}}' , ~~~~~ v_{z}^{2} = b^{-\frac{z+1}{z}} {v_{z}^{2}}' , ~~~~~ m_{z}^{2} = b^{-2} {m_{z}^{2}}' . \eqa
%
%
%
Interaction parameters are transformed as follows \bqa && \Bigl(\frac{g_{1} g_{z}}{N_{\sigma} m_{1}^{2}}\Bigr) = b^{-\frac{1-z}{2z}} \Bigl(\frac{g_{1} g_{z}}{N_{\sigma} m_{1}^{2}}\Bigr)' , ~~~~~ g_{2} = b^{- \frac{3-z}{4z}} g_{2}' , ~~~~~ e_{f} = e_{f}' , ~~~~~ e_{z} = b^{- \frac{1-z}{z} } e_{z}' . \eqa

If we consider the case of $1 < z < 3$, both the spinon-holon coupling constant $\Bigl(\frac{g_{1} g_{z}}{N_{\sigma} m_{1}^{2}}\Bigr)$ and spinon-gauge coupling constant $e_{z}$ turn out to be relevant while the holon-gauge coupling constant $e_{f}$ is marginal. The coupling constant $g_{2}$ is irrelevant, implying that the Hertz-Moriya-Millis theory should be modified for this antiferromagnetic quantum criticality.
As a result, we obtain an effective field theory \bqa && Z = \int D f_{s\sigma} D z_{\sigma} D a \exp\Bigl[ - \int_{0}^{\beta} d \tau \int_{-\infty}^{\infty} d x \int_{-\infty}^{\infty} d y \Bigl\{ f_{s\sigma}^{\dagger}(\tau,x,y) \Bigl(- i \frac{c_{2}}{N_{\sigma}} (- \partial_{\tau}^{2})^{\frac{z-1}{2z}} - i s v_{F} \partial_{x} - \frac{v_{F}}{2\gamma} \partial_{y}^{2} \Bigr) f_{s\sigma}(\tau,x,y) \nn && + a(\tau,x,y) \gamma \frac{\sqrt{-\partial_{\tau}^{2}}}{\sqrt{-\partial_{y}^{2}}} a(\tau,x,y) - \frac{e_{f}}{\sqrt{N_{\sigma}}} s a(\tau,x,y) v_{F} \sigma f_{s\sigma}^{\dagger}(\tau,x,y) f_{s\sigma}(\tau,x,y) + |\partial_{\tau} z_{\sigma}(\tau,x,y)|^{2} \nn && - \frac{g_{1} g_{z}}{N_{\sigma} m_{1}^{2}} \sigma f_{s\sigma}^{\dagger} f_{s\sigma} (z_{\sigma'}^{\dagger} \partial_{\tau} z_{\sigma'}) - i \frac{e_{z}}{\sqrt{N_{\sigma}}} a [ z_{\sigma}^{\dagger} (\partial_{x} z_{\sigma}) - (\partial_{x} z_{\sigma}^{\dagger}) z_{\sigma}] \Bigr\} \Bigr] , \eqa not a fixed-point theory.

On the other hand, if we take the scale transformation as follows \bqa && \omega = b^{-1} \omega' , ~~~~~ k_{\parallel} = b^{-\frac{z-1}{z}} k_{\parallel}' , ~~~~~ k_{\perp} = k_{\perp}' , \eqa we obtain \bqa && f_{s\sigma}(\omega,k_{\parallel},k_{\perp}) = b^{\frac{3z-2}{2z}} f_{s\sigma}'(\omega',k_{\parallel}',k_{\perp}') , ~~~~~ \phi_{2}(\omega,k_{\parallel},k_{\perp}) = b^{\frac{2z+1}{2z}} \phi_{2}'(\omega',k_{\parallel}',k_{\perp}') , \nn && a(\omega,k_{\parallel},k_{\perp}) = b^{\frac{3z-1}{2z}} a'(\omega',k_{\parallel}',k_{\perp}') , ~~~~~ z_{\sigma}(\omega,k_{\parallel},k_{\perp}) = b^{\frac{4z-1}{2z}} z_{\sigma}'(\omega',k_{\parallel}',k_{\perp}') . \eqa Then, all interactions turn out to be irrelevant under this scale transformation as follows
%
%
%
%
%
\bqa && g_{2} = b^{- \frac{1}{2z}} g_{2}' , ~~~~~ e_{f} = b^{- \frac{2z-1}{2z} } e_{f}' , ~~~~~ e_{z} = b^{- \frac{z+3}{2z} } e_{z}' \eqa except for the spinon-holon coupling constant, which remains to be still marginal \bqa && \Bigl(\frac{g_{1} g_{z}}{N_{\sigma} m_{1}^{2}}\Bigr) = \Bigl(\frac{g_{1} g_{z}}{N_{\sigma} m_{1}^{2}}\Bigr)' . \eqa
%
%
The critical field theory for this fixed point is given by \bqa && Z = \int D f_{s\sigma} D z_{\sigma} \exp\Bigl[ - \int_{0}^{\beta} d \tau \int_{-\infty}^{\infty} d x \Bigl\{ f_{s\sigma}^{\dagger}(\tau,x) \Bigl(- i \frac{c_{2}}{N_{\sigma}} (- \partial_{\tau}^{2})^{\frac{z-1}{2z}} - i s v_{F} \partial_{x} \Bigr) f_{s\sigma}(\tau,x) \nn && + |\partial_{\tau} z_{\sigma}(\tau,x)|^{2} - \frac{g_{1} g_{z}}{N_{\sigma} m_{1}^{2}} \sigma f_{s\sigma}^{\dagger}(\tau,x) f_{s\sigma}(\tau,x) [z_{\sigma'}^{\dagger}(\tau,x) \partial_{\tau} z_{\sigma'}(\tau,x)] \Bigr\} \Bigr] , \eqa which implies the robustness of local quantum criticality.

\section{Summary}

In this study we revisited an antiferromagnetic quantum phase transition from a Fermi-liquid state. In particular, we focused on the ordering wave vector of $\bm{Q} = 2 \bm{k}_{F}$, where the Fermi velocity of one patch is parallel to that of the other, where both patches are connected by the nesting vector. In this situation the scale transformation for the longitudinal momentum orthogonal to the Fermi surface differs from that for the transverse momentum along the Fermi surface. An idea was to extract out dynamics of directional spin fluctuations explicitly from the Hertz-Moriya-Millis theory. Taking the strong coupling approach which diagonalizes the spin-fermion coupling term in the Hertz-Moriya-Millis theory, we could deal with both amplitude and directional fluctuations of the antiferromagnetic order parameter on equal footing. As a result, we constructed an effective field theory Eq. (21) in the two-patch construction. Recall Eq. (14) for the lattice construction. Introducing quantum corrections into the U(1) slave spin-rotor effective field theory in the one-loop level, we found a renormalized field theory Eq. (24) for the antiferromagnetic phase transition with $\bm{Q} = 2 \bm{k}_{F}$, which describes dynamics of renormalized electrons, $z = 2$ critical longitudinal spin (antiferromagnetic amplitude) fluctuations, $z = 1$ transverse (directional) spin fluctuations, $z = 3$ U(1) gauge fluctuations, and their interactions, where $z$ is the dynamical critical exponent. Considering several ways of scale transformations, we could find a critical field theory in terms of renormalized electrons and transverse spin fluctuations, where only their interactions turn out to be marginal while all others become irrelevant. A characteristic feature was that such transverse spin fluctuations become localized, which means that their propagating velocity is renormalized to vanish at this fixed point. Emergence of localized magnetic moments is an interesting novel feature of this study, proposing local quantum criticality for the antiferromagnetic quantum phase transition with $\bm{Q} = 2 \bm{k}_{F}$.

\section*{Acknowledgement}

This study was supported by the Ministry of Education, Science, and Technology (No. 2012R1A1B3000550) of the National Research Foundation of Korea (NRF) and by TJ Park Science Fellowship of the POSCO TJ Park Foundation.

\end{document}